\documentclass[
preprint,
superscriptaddress,
nofootinbib,
 amsmath,amssymb,
 prb,
{article}
]{revtex4-2}

\usepackage{comment}
\usepackage{amsfonts}
\usepackage{graphicx}
\usepackage{physics}
\usepackage{color}
\usepackage{amsthm}
\usepackage{tikz}
\usepackage{tikz-cd}
\usepackage{tikzit}
\usepackage{CJK}

\usepackage{tikz,lipsum,lmodern}
\usepackage[most]{tcolorbox}

\usepackage{relsize}
\usepackage{hyperref}
\usepackage{caption}
\usepackage{subcaption}
\usepackage{dutchcal}
\usepackage{algpseudocode,algorithm}
\usepackage{mathtools}

\tikzstyle{red node}=[fill=red, tikzit category=nodes, shape=circle, draw=black]
\tikzstyle{blue node}=[fill=blue, shape=circle, draw=black, tikzit category=nodes]
\tikzstyle{green node}=[tikzit fill=green, fill=green, shape=circle, draw=black, tikzit category=nodes]
\tikzstyle{yellow square}=[draw=black, fill=yellow, shape=rectangle]
\tikzstyle{blue node 2}=[fill={rgb,255: red,128; green,0; blue,128}, draw=black, shape=circle, tikzit fill=blue]
\tikzstyle{black node}=[fill=black, draw=black, shape=circle]
\tikzstyle{Rectangle}=[fill=none, draw=black, shape=rectangle]

\tikzstyle{dashed edge}=[<->, dashed]
\tikzstyle{blue pointer}=[->, draw=blue]
\tikzstyle{black}=[->]
\tikzstyle{new edge style 0}=[-]

\tikzstyle{morphism}=[fill=white, draw=black, shape=rectangle]
\tikzstyle{medium box}=[fill=white, draw=black, shape=rectangle, minimum width=0.8cm, minimum height=0.9cm]
\tikzstyle{large morphism}=[fill=white, draw=black, shape=rectangle, minimum width=1.7cm, minimum height=1cm]
\tikzstyle{bn}=[fill=black, draw=black, shape=circle, inner sep=1.5pt]
\tikzstyle{state}=[fill=white, draw=black, regular polygon, regular polygon sides=3, minimum width=0.8cm, shape border rotate=180, inner sep=0pt]
\tikzstyle{medium state}=[fill=white, draw=black, regular polygon, regular polygon sides=3, minimum width=1.3cm, inner sep=0pt, shape border rotate=180]
\tikzstyle{large state}=[fill=white, draw=black, regular polygon, regular polygon sides=3, minimum width=2.2cm, shape border rotate=180, inner sep=0pt]
\tikzstyle{wn}=[fill=white, draw=black, shape=circle, inner sep=1.5pt]
\tikzstyle{wide state}=[fill=white, draw=black, shape=isosceles triangle, minimum width=0.8cm, shape border rotate=270, inner sep=1.4pt, minimum height=0.5cm, isosceles triangle apex angle=80]
\tikzstyle{evalold}=[fill=white, draw=black, shape=isosceles triangle, minimum width=1.4cm, shape border rotate=90, inner sep=1.4pt, minimum height=0.4cm, isosceles triangle apex angle=110]
\tikzstyle{eval}=[fill=white, draw=black, shape=rectangle, minimum width=1.4cm, minimum height=0.55cm, inner sep=1.4pt, font={$\eval$}]

\tikzstyle{arrow}=[->]
\tikzstyle{dashed box}=[-, dashed]
\tikzstyle{mapsto}=[{|->}]
\tikzstyle{double wire}=[-, double]
\tikzstyle{protected}=[-, preaction={{ultra thick,white,draw}}]
\tikzstyle{ambient fill}=[-, draw=none, fill={rgb,255: red,245; green,220; blue,255}, tikzit draw={rgb,255: red,210; green,130; blue,255}]

\usetikzlibrary{positioning}

\usepackage{todonotes}

\usepackage{multirow}
\usepackage{dcolumn}

\newcommand{\beq}{\begin{equation}\begin{aligned}}
\newcommand{\eeq}{\end{aligned}\end{equation}}

\newcommand{\Z}{{\mathbb Z}}

\begin{document}
\begin{CJK*}{UTF8}{}

\title{Measuring Topological Field Theories: Lattice Models and Field-Theoretic Description}

\author{Yabo Li (\CJKfamily{gbsn}李雅博)}
\email{li.yabo@stonybrook.edu}
\affiliation{C. N. Yang Institute for Theoretical Physics, State University of New York at Stony Brook, New York 11794-3840, USA}
\affiliation{Department of Physics and Astronomy, State University of New York at Stony Brook, New York 11794-3840, USA}
\author{Mikhail Litvinov}
\affiliation{C. N. Yang Institute for Theoretical Physics, State University of New York at Stony Brook, New York 11794-3840, USA}
\affiliation{Department of Physics and Astronomy, State University of New York at Stony Brook, New York 11794-3840, USA}
\author{Tzu-Chieh Wei (\CJKfamily{bsmi}魏子傑)}
\affiliation{C. N. Yang Institute for Theoretical Physics, State University of New York at Stony Brook, New York 11794-3840, USA}
\affiliation{Department of Physics and Astronomy, State University of New York at Stony Brook, New York 11794-3840, USA}

\begin{abstract}
Recent years have witnessed a surge of interest in performing measurements within topological phases of matter, e.g., symmetry-protected topological (SPT) phases and topological orders. Notably, measurements of certain SPT states have been known to be related to Kramers-Wannier duality and Jordan-Wigner transformations, giving rise to long-range entangled states and invertible phases, such as the Kitaev chain. Moreover, measurements of topologically ordered states correspond to charge condensations. In this work, we present a field-theoretic framework for describing measurements within topological field theories. We employ various lattice models as examples to illustrate the outcomes of measuring local symmetry operators within topological phases, demonstrating their agreement with the predictions from field-theoretic descriptions. We demonstrate that these measurements can lead to SPT, spontaneous symmetry-breaking, and topologically ordered phases. Specifically, when there is emergent symmetry after measurement, the remaining symmetry and emergent symmetry will have a mixed anomaly, which leads to long-ranged entanglement.

\end{abstract}
\maketitle
\newpage
\end{CJK*}

\tableofcontents

\section{Introduction}

In recent decades, there has been significant progress in exploring topological phases of matter, which is not described by the Landau theory of symmetry spontaneously breaking.  Notable examples of these new phases include the intrinsic topological orders which have degeneracy on topological non-trivial closed manifolds~\cite{wen1990topological,wenniu1989ground,kitaev2003fault,levin2005string,kitaev2006anyons,wen2016theory}, and the symmetry protected topological (SPT) orders, which have a unique gapped ground state in closed manifolds, but exhibit a rich physics due to the symmetry anomalies on the boundary~\cite{chen2011classification,turner2011topological,schuch2011classifying,lu2012theory,chen2013symmetry,chen2014symmetry,senthil2015symmetry}. While the topological phases can, in principle, be described by topological field theories~\cite{atiyah1988topological,witten1989quantum}, numerous topological phases have been constructed as solvable models on lattices~\cite{dijkgraaf1990topological,raussendorf2002one,kitaev2003fault,levin2005string,walker20123+,hu2013twisted,cheng2017exactly} and these have in turn given rise to much insight of the phases.

In the past decade, there have been extensive generalizations of SPT phases with higher-form symmetries~\cite{kapustin2017higher,yoshida2016topological,benini20192,tsui2020lattice,jian2021physics,hsin2022exotic}.
The topological actions of generalized bosonic SPT phases can be constructed and classified using cobordism~\cite{kapustin2014symmetry,kapustin2015fermionic,wan2018higher}. It has been recently argued that the Higgs phase should be considered as such a generalized SPT phase~\cite{verresen2022higgs}. The concept of higher-form symmetry also allows for a fresh perspective on intrinsic topological order~\cite{kapustin2014coupling,gaiotto2015generalized,cordova2019anomaly}. For example, in this context, the deconfined phase of the $\mathbb{Z}_2$ gauge theory for $d\geq2$ that is characterized by the toric code topological order spontaneously breaks a $(d-1)$-form symmetry. 

SPT states have also been recognized as resource states for measurement-based quantum computing~\cite{briegel2009measurement,wei2018quantum,raussendorf2002one,miyake2010quantum,else2012symmetry,stephen2017computational,raussendorf2017symmetry,wei2017universal,raussendorf2019computationally}.
More recently, there has been a mounting interest in performing measurements within topological phases.
Creating an SPT state or a topological order state on a quantum computer is a tricky business, and ideally, one wants to have a finite-depth circuit when working with real devices to avoid noise from piling up. 
Of course, one cannot achieve this when preparing an SPT state with unitary operations without breaking the symmetry, and the ground state of the topological order requires an non-constant-depth circuit. To overcome this problem, one can try to introduce measurement (a non-unitary operation) into the mix.
 In specific scenarios, conducting measurements on SPT states can lead to the creation of topologically ordered states. For instance, the ground state of the toric code can be efficiently prepared by measuring (and correcting) part of qubits in the ($2+1$)d cluster state~\cite{raussendorf2005long}. There, measurements can be seen as a way to transform from the Higgs phase (represented by the cluster state) to the deconfined phase (represented by the toric code)~\cite{verresen2022higgs}.
 A roadmap has been developed for creating a large class of long-range entangled states from 0-form SPT states with finite-depth operations including local unitary operations, measurements, feedforward, and corrections~\cite{piroli2021quantum, lu2022measurement, bravyi2022adaptive, ashkenazi2022duality,tantivasadakarn2023hierarchy,li2023symmetry}.

In general, measuring the higher-form symmetry operators in topologically ordered states can be seen as condensing charged particles. It has been demonstrated that through particle condensation in the bulk of some topological order, certain SPT phases and topological orders can be realized~\cite{jiang2017anyon,ellison2022pauli}. Moreover, by condensing particles on a sub-manifold, we can construct gapped boundaries and condensation defects of a topological order~\cite{kapustin2011topological,yoshida2017gapped,roumpedakis2023higher}. It is worth noting that the anyon condensation operations in topological orders play a crucial role in facilitating fault-tolerant quantum computing~\cite{kesselring2022anyon}.

Despite the fruitful results of performing measurements in topological phases, a general framework that describes measurements in topological phases is still missing. Building on a prior work~\cite{tantivasadakarn2021long}, in which a class of SPT phases is argued to give rise to anomalous long-range entangled states upon measurements, here we use a field-theoretic formalism and provide a systematic approach to describe measurements. We employ various lattice models as examples to illustrate the outcomes of measuring local symmetry operators within topological phases, demonstrating their agreement with the predictions from field-theoretic descriptions. Specifically, we address the following:
\begin{enumerate}
    \item Measuring a subgroup of local symmetry operators in a generalized SPT state may result in either an SPT state or a long-range entangled state, depending on  the emergent symmetry after measurement. (See Sec.~\ref{sec:measure_0_form} and Sec.~\ref{sec:measure_generalized}.)
    \\
    \item The long-range entangled state resulting from measuring an SPT state exhibits a mixed anomaly between the remaining symmetry and the emergent symmetry. This mixed anomaly allows us to infer the phase of the measured state, which can be a spontaneously symmetry-breaking (SSB) phase, a topological order, or a symmetry-enriched topological (SET) order. (See Sec.~\ref{sec:measure_0_form} and Sec.~\ref{sec:measure_generalized}.)
    \\
    \item Measuring a subgroup of local symmetry operators in a topologically ordered state may lead to a different topologically ordered state, corresponding to the outcome of particle condensation in the original order. (See Sec.~\ref{sec:measure top order}.)
\end{enumerate}

This work is organized as follows. In Sec.~\ref{sec:teasers}, as a motivation, we use cluster states to illustrate how measured states from SPTs can exhibit mixed anomalies between remaining and emergent symmetries, leading to SSB or topologically ordered phases. In Sec.~\ref{sec:theoretic}, we review the topological actions of SPT phases and demonstrate how SPT states can be derived from these actions, then we present a general procedure for obtaining the action for the measured phases. We illustrate this procedure with examples in Sections~\ref{sec:measure_0_form}, ~\ref{sec:measure_generalized}, and ~\ref{sec:measure top order}, where we discuss performing measurements in 0-form SPTs, generalized SPTs, and topological orders. We use both the field-theoretic descriptions and the lattice model to analyze the phases of the measured states and the emergent anomalies. In Sec.~\ref{sec:conclusion}, we make some concluding remarks. App.~\ref{appendix:spt state}, \ref{appendix:Z2_sub_measure1} and \ref{appendix:Z2_cub_cont} provide detailed calculations of constructing $\mathbb{Z}_2^3$ type-3 SPT lattice model and the analysis of the measured states.

\section{Teaser: from cluster SPT}
\label{sec:teasers}
In this section, we consider a couple of examples to showcase how after the measurement a topologically ordered state emerges with the quantum symmetry.

We first consider the $\mathbb{Z}_2\times\mathbb{Z}_2$ SPT state on a $1$d lattice.
The Hamiltonian of the model is given by
\beq
    H=-\sum_{i\in\mathbb{Z}} Z_{i-1} X_i Z_{i+1},
\eeq
where $X$ and $Z$ are the Pauli operators for qubits. The ground state of this Hamiltonian is the so-called cluster state. If we perform a projective measurement on all the odd sites by $X$ and post-select the measurement outcome to be $X_{2k+1}=1$ for any $k\in \mathbb{Z}$. The stabilizers for the measured state are $X_{2k+1}$, $Z_{2k}Z_{2k+2}$ and $\prod_{k}X_{2k}$. Ignoring the disentangled odd sites, the measured state is essentially GHZ state $\ket{0 \dots 0}+\ket{1\dots1}$ formed by even sites spins. The resultant parent Hamiltonian is the sum of the Z stabilizers,
\beq
    H'=-\sum_{j\in 2\mathbb{Z}} Z_{j-2}Z_{j}.
\eeq

The measured state has an anomalous $\mathbb{Z}_2\times\mathbb{Z}_2^{(1)}$ symmetry~\cite{pace2023emergent}.
The ordinary $\mathbb{Z}_2$ symmetry $\prod_j X_j$ is supported on the entire $1$d lattice, while the 1-form $\mathbb{Z}_2$ symmetry $Z_{i}Z_{j}$ is supported on the boundary of any $1$d segment.
There are two ways to characterize the anomaly on the lattice.
The first way is to note that the localized symmetry operators (also called patch symmetry transformation in Ref.~\cite{ji2020categorical}) $\prod_{k\in s}X_k$ for a segment $s$, and $Z_j$ anti-commute when $j\in s$. The second way is to make a virtue of the fact that an anomalous state can live on the boundary of a corresponding SPT state.
In this case, the SPT is exactly a ($2+1$)d cluster state as pointed out in Ref.~\cite{verresen2022higgs}, and will be shown below.

\begin{figure}[h]
\centering
\includegraphics[width=0.8\linewidth]{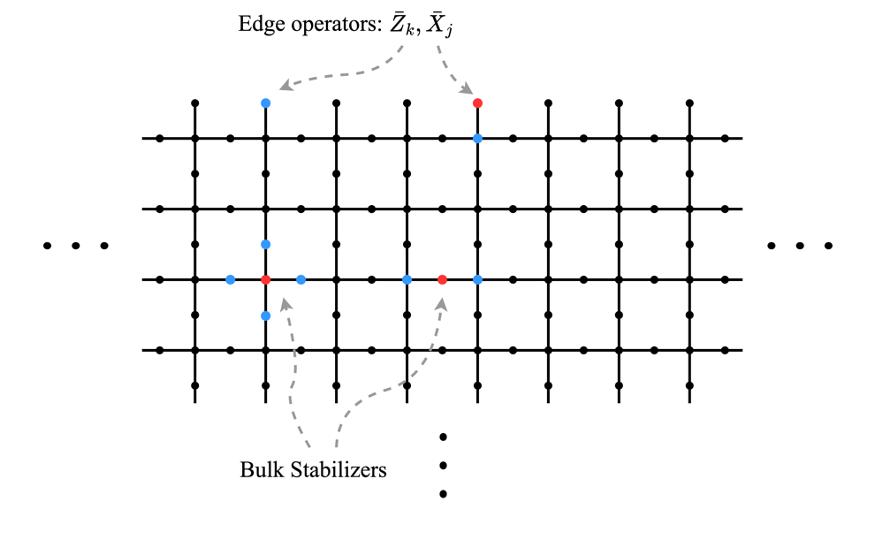}
\caption{\label{fig:2d_cluster} The ($2+1$)d $\mathbb{Z}_2\times\mathbb{Z}_2^{(1)}$ cluster state and its rough boundary on the square lattice, where qubits live on both sites and links. Pauli $X$ operators are denoted as red, and Pauli $Z$ operators are denoted as blue. There are two types of bulk stabilizers corresponding to $\mathbb{Z}_2$ 0-form and $\mathbb{Z}_2$ 1-form symmetries. The symmetry operators on the boundary are composed of the edge operators $\bar{Z}$ and $\bar{X}$. The boundary symmetry operators are exactly the same as the GHZ state from measuring ($1+1$)d cluster state (up to a Hadamard).}
\end{figure}

This ($2+1$)d cluster state, with both 0-form and 1-form symmetry, can be put on a square lattice. The Hamiltonian is given by the sum of bulk stabilizers,
\beq
    H=-\sum_v X_v \prod_{e\supset v}Z_e - \sum_e X_e \prod_{v\subset e}Z_v.
\eeq
The global symmetry operators on a square lattice without boundary are
$\prod_v X_v,\ \prod_{e\in c}X_e$,
for any closed loop $c$. And the ground state is given by $\ket{SPT}=\mathcal{U}\ket{+}^{\otimes N}$, where $\mathcal{U}=\prod_{v\subset e}CZ_{v,e}$, and $N$ is the number of all qubits on edges and vertices. On a square lattice with a boundary as in Fig.~\ref{fig:2d_cluster}, there are dangling degrees of freedom at the edge. The edge operators are given by
$\bar{X}_j:=\mathcal{U}X_j\mathcal{U}^{\dagger},\ \bar{Z}_j:=\mathcal{U}Z_j\mathcal{U}^{\dagger}$, where index $j$ is the label on the $1$d boundary~\cite{han2023topological}. When acting on the SPT state with boundary, the symmetry operators can be decomposed into edge operators,
\beq
    \prod_v X_v \ket{SPT} &= \prod_j \bar{Z}_j\ket{SPT},\\
    \prod_{e\in c} X_e \ket{SPT} &= \bar{X}_i \bar{X}_j\ket{SPT},
\eeq
for any string $c$ that ends at boundary edge $i$ and $j$. Therefore $\prod_j \bar{Z}_j$ and $\bar{X}_i \bar{X}_j$ are the symmetry operators on the boundary theory, see also Fig.~\ref{fig:2d_cluster}. 

After explaining that the 1d GHZ state can live on the boundary of this SPT state, we now consider it on a closed $2$d lattice.  If we perform a projective measurement on all the vertices qubits by $X$ and post-select $X_v=1$ for all $v$, ignoring the disentangled vertices spins, the stabilizers for the measured state are $\prod_{e\supset v}Z_e$ and $\prod_{e\in p}X_{e}$ for any vertex $v$ and any plaquette $p$.
Those stabilizers are exactly the star and plaquette operators for a toric code on the square lattice. 
Therefore after measurement, we obtain a ground state of the toric code model.
The toric code model (described by a deconfined $\mathbb{Z}_2$ gauge theory) is known to have an anomalous $\mathbb{Z}_2^{(1)}\times\mathbb{Z}_2^{(1)}$ symmetry. We can characterize the anomaly on the lattice similar to the case in $1$d, on the one hand, by the fact that the symmetry operators of the 1-form symmetries locally anti-commute. On the other hand, we also note that toric code can be put on the boundary of a ($3+1$)d SPT
~\cite{raussendorf2005long, yoshida2016topological}.

If we instead, measure all the edge qubits with observable $X$ and post-select $X_e=1$ for all $e$, the stabilizers for the measured state are $Z_i Z_j$ for any vertices $i$ and $j$, and $\prod_k X_k$.
Therefore, after measurement, we obtain a GHZ state. By the similar argument from above, the GHZ state has an anomalous $\mathbb{Z}_2\times\mathbb{Z}_2^{(2)}$ symmetry, and can be put on the boundary of a ($3+1$)d SPT~\cite{pace2023emergent}.

From above, we show that measuring an SPT state can lead to a state with emergent anomaly, which corresponds to a symmetry spontaneously broken state, or topologically ordered state.
The analysis is based on specific lattice constructions, but due to the topological nature of the SPT phases, we expect the results to hold for any lattice construction of the same SPT phase.
In order to characterize the phases of measured states and the potential emergent anomalies in more general cases, we aim in this work to construct a field-theoretic framework for studying the phases 
and topological properties after measurements.
In the ensuing sections, we will demonstrate the application of this framework in measuring SPTs through multiple examples.
Furthermore, we will illustrate the generality of the framework by showcasing its application in measuring topological orders.

\section{Theoretical framework}
\label{sec:theoretic}
\subsection{SPT topological actions}
\label{sec:top action}
In this subsection, we briefly review the topological actions of SPT order and its relation with the fixed-point wavefunction~\cite{wang2015field,tsui2020lattice,chen2023higher}.

For a $G^{(p)}$ SPT state, with a $p$-form abelian symmetry group $G$, we can couple the symmetry with a background gauge field $A$, which is a $G$-valued $(p+1)$-cocycle. The topological action is given by
\beq
    S^{\text{top}}[M,A]=2\pi i \int_{M} \mathcal{L}[A],
\eeq
where the Lagrangian $\mathcal{L}[A]$ being a $(d+1)$-cocycle, satisfies the cocycle condition,
\beq
    \delta\mathcal{L}[A]\overset{1}{=} 0.
\eeq

If two Lagrangians differ by a coboundary,
\beq
    \mathcal{L}'=\mathcal{L} + \delta \Theta,
\eeq
where $\Theta$ could be any $d$-cochain, then the actions only differ by a boundary term. When the spacetime manifold is closed, the two actions are exactly the same. Therefore we call two Lagrangians differing by only a coboundary as equivalent. Distinct inequivalent Lagrangians characterize distinct bosonic $G^{(p)}$ SPT orders. This means that SPT orders should be given by some cohomology classes. Indeed, for a $0$-form bosonic symmetry $G$, the SPT orders in $d+1$ dimensional spacetime are characterized by cohomology group $H^{d+1}(BG,U(1))$. We show in Table~\ref{table:cocycle} some of the correspondences between the group cocycles $\omega\in H^{d+1}(G,U(1))\simeq H^{d+1}(BG,U(1))$ and the SPT partition functions, which will be used later. For a general higher-group $\mathbb{G}$, the SPT orders in $d+1$ dimensional spacetime are characterized by cohomology group $H^{d+1}(B\mathbb{G},U(1))$.

\begin{table*}[t]
\centering
\begin{tabular}{ |m{0.05\linewidth}|m{0.09\linewidth}|m{0.15\textwidth}|m{0.29\textwidth}|m{0.4\textwidth}| } 
\hline
$d+1$ & \multicolumn{1}{c|}{$G$} & \multicolumn{1}{c|}{$H^{d+1}(G,U(1))$} & \multicolumn{1}{c|}{partition function $Z$} & \multicolumn{1}{c|}{group cocycle $\omega(a,b,...)$}  \\
\hline
1+1 & \multirow{4}{*}{$\mathbb{Z}_m\times\mathbb{Z}_n$} & $\mathbb{Z}_{gcd(m,n)}$ & $\exp(\int\frac{2\pi i k}{gcd(m,n)}A_1 \cup A_2)$  & $\exp(\frac{2\pi i k}{gcd(m,n)}a_1 b_2)$\\ 
\cline{1-1}
\cline{3-5}
\multirow{3}{*}{2+1} & & 
$\mathbb{Z}_{gcd(m,n)}$ & $\exp\left(\int\frac{2\pi i k}{mn}A_1\cup \delta A_2\right)$ & $\exp\left(\frac{2\pi i k}{mn}a_1 (b_2+c_2-[b_2+c_2]_n)\right)$ \\
 & & $\times\mathbb{Z}_m$ & $\exp\left(\int\frac{2\pi i k'}{m^2}A_1 \cup \delta A_1\right)$ & 
$\exp\left(\frac{2\pi i k'}{m^2}a_1 (b_1+c_1-[b_1+c_1]_m)\right)$ \\
 & & $\times\mathbb{Z}_n$ & $\exp\left(\int\frac{2\pi i k''}{n^2}A_2 \cup \delta A_2\right)$ & $\exp\left(\frac{2\pi i k''}{n^2}a_2 (b_2+c_2-[b_2+c_2]_n)\right)$ \\ 
\hline
\multirow{3}{*}{2+1} & \multirow{3}{*}{$\mathbb{Z}_2^3$} & $\mathbb{Z}_2^3$ (type-1) & $\exp\left(\int\sum_j\frac{2\pi i k_j}{4}A_j\cup \delta A_j\right)$ & $\exp\left(\sum_j\frac{2\pi i k_j}{4}a_j(b_j+c_j-[b_j+c_j]_2)\right)$ \\ 

 &  & $\times\mathbb{Z}_2^3$ (type-2) & $\exp\left(\int\sum_{i< j}\frac{2\pi i k_{ij}}{4}A_i\cup \delta A_j\right)$ & $\exp\left(\sum_{i< j}\frac{2\pi i k_{ij}}{4}a_i(b_j+c_j-[b_j+c_j]_2)\right)$ \\ 

 &  & $\times\mathbb{Z}_2$ (type-3) & $\exp(\int\frac{2\pi i k'}{2}A_1\cup A_2 \cup A_3)$ & $\exp(\frac{2\pi i k'}{2}a_1 b_2 c_3)$ \\ 
\hline
\end{tabular}
\caption{Some correspondences between SPT partition functions and group cocycles $\omega$ for 0-form symmetries in ($1+1$)d and ($2+1$)d; see Ref.~\cite{wang2015field}. The parameters $k$, $k'$, $k''$, $k_j$, and $k_{ij}$ take value in the corresponding ranges.}
\label{table:cocycle}
\end{table*}
We can also find a physical action that produces this SPT state as a ground state. 
Usually, such constructions are decorated  domain wall constructions~\cite{chen2014symmetry}. We take a different route and use a formalism developed in Ref.~\cite{tsui2020lattice}.
To realize the SPT states characterized by a topological action, we replace the $\Z_n$-valued gauge field $A$ in $\mathcal{L}[A]$ by $[\delta\phi]_n$, where $[x]_n:= x\ \text{mod}\ n$, and the cochain $\phi$ is the physical field of the SPT states.
In what follows we call the action with physical fields \emph{the physical action}.
The global symmetry $\phi\rightarrow\phi+c$ of $\mathcal{L}[\delta\phi]$ for any cocycle $c$, is inherited from the gauge invariance of $\mathcal{L}[A]$. The partition function of the SPT order, after coupling to the background gauge field $A$ and summing all configurations of the physical field $\phi$, reproduces the SPT topological action as a low-energy action~\cite{kong2014braided},
\beq
    Z[M,A]=\sum_{\phi\in C^p(M,G)}e^{2\pi i\int_M \mathcal{L}[\delta\phi+A]} \propto e^{S_{\text{top}}(A)}.
\eeq

In general, the path integral of a quantum field theory on a manifold $M$ defines an quantum state on its boundary $\partial M$, $\ket{\psi_M}=\sum_{\phi|_{\partial M} =\phi_b}e^{S[\phi]}\ket{\phi_b}$. The Lagrangian $\mathcal{L}[\delta\phi]$ of a SPT phase is a coboundary, $\mathcal{L}[\delta\phi]=\delta\omega[\phi]$. Therefore, the SPT state defined on a manifold $M$ does not depend on the bulk of $M$ at all, and can be well-defined just on the boundary. The SPT state on a closed spatial manifold $\partial M$ is given by 
\beq
    \ket{SPT}=\sum_{\phi\in C^p(\partial M,G)}e^{2\pi i\int_{\partial M} \omega[\phi]}\ket{\phi},
\eeq
for which we have suppressed a proper normalization of the wavefunction. 
This relation establishes a connection between a topological action and an SPT state explicitly.

For pedagogical purposes, let us explain the simplest possible example: the cluster state in ($1+1$)d. Of course, we expect that we get the usual ground state of  $H =- \sum_i Z_{i-1}X_i Z_{i+1}$.
The topological action of this model is given by 
\begin{equation}
    S^{\text{top}} = 2 \pi i  \int_M \frac{1}{2} A \cup B,
\end{equation}
where $A$, $B$ are $\mathbb{Z}_2$-valued $1$-cocycle. After the substitution $A\rightarrow [\delta\phi_a]_2,\ B\rightarrow [\delta\phi_b]_2$, where $[x]_n:= x\ \text{mod}\ n$, the physical action of the SPT order is given by
\beq
    S=2\pi i \int_M \frac{1}{2}\delta\phi_a \cup \delta\phi_b=2\pi i \int_M \delta\left(\frac{1}{2}\phi_a\cup \delta\phi_b\right), 
\label{eq:before_measure}
\eeq
where $\phi_a,\phi_b \in C^0(M,\mathbb{Z}_2)$ and we have used $\delta^2\phi_b=0$. The SPT state on the spatial manifold $\partial M$ is then given by
\beq
    \ket{SPT}=\sum_{\phi_a,\phi_b}e^{2\pi i \int_{\partial M}\frac{1}{2}\phi_a\cup \delta\phi_b}\ket{\phi_a,\phi_b}.
\eeq

\begin{figure}[h]
\includegraphics[width=0.6\linewidth]{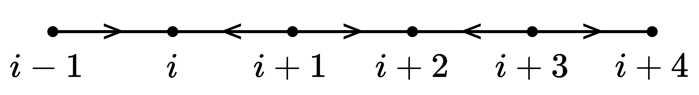}
\caption{\label{fig:1d_lattice} Triangulation of the $1$d lattice.}
\end{figure}

We triangulate the $1$d manifold as in Fig.~\ref{fig:1d_lattice}, where $i$ is even. On each site, we have one qubit of type $a$ and one qubit of type $b$.
With this triangulation, let us consider this action on edge $(i,i-1)$ and edge $(i,i+1)$  (both being 1-simplices):
\begin{equation}
\begin{split}
    \phi_a \cup \delta\phi_b(\sigma_1+\sigma_2) &= \phi_a(i)\left(\phi_b(i)-\phi_b(i-1)\right) \\
    &+ \phi_a(i)(\phi_b(i)-\phi_b(i+1)) \overset{2}{=}\phi_a(i)(\phi_b(i-1)+\phi_b(i+1)),
\end{split}
\end{equation}
where, on the last line, we have used the $\Z_2$ property of one chain.
So, we see that the $b$ qubits on even sites and $a$ qubits on odd sites do not contribute; hence, they are decoupled (i.e., unentangled with the rest). In fact, it's just a decoration of $\mathbb{Z}_2$ domain wall. 
To obtain a familiar expression, we choose physical fields $\phi_a$ and $\phi_b$ to be qubits in the $Z$-basis.
If we omit all disentangled qubits, there is only one qubit per site. Therefore, we can omit the qubit types and only use a site number to denote each qubit, then
\beq
    \ket{SPT}&=\sum_{\phi} e^{\sum_{i\ \text{even}}\phi(i)(\phi(i-1)+\phi(i+1))}\ket{\{\phi(i)\}}\\
    &=(-1)^{\sum_{i\ \text{even}}\frac{1-Z^{(a)}_i}{2}\frac{1-Z^{(b)}_{i-1}}{2}+\frac{1-Z^{(a)}_i}{2}\frac{1-Z^{(b)}_{i+1}}{2}}\ket{+}^{\otimes n}\\
    &=\prod_{i\ \text{even}}CZ_{i,i-1}CZ_{i,i+1}\ket{+}^{\otimes n},
\eeq
where in going from the first to the second line we used $\phi(i)=\frac{1}{2}(1-Z_i)$.
This is the standard form of the $1$d cluster state in the literature. The stabilizers of the state are $Z_{i-1}X_i Z_{i+1}$~\footnote{If we choose another branching structure instead of the colorable one we use, for each edge that swaps its orientation, the local wavefunction amplitude on this edge can be written down in the colorable branching structure as $\delta\phi_b\cup \phi_a(i,i+1)$. Therefore the difference in the wavefunction amplitude is $(-1)^{\delta\phi_b\cup \phi_a-\phi_a\cup \delta\phi_b}=(-1)^{\delta\phi_a\cup_1 \delta\phi_b(i,i+1)}$. This is a local operator symmetric under $\mathbb{Z}_2\times\mathbb{Z}_2$, because it is invariant under $\phi_a\rightarrow\phi_a+c_1$ and $\phi_a\rightarrow\phi_a+c_1$. Therefore, we conclude that different branching structures of the lattice will give different SPT states related by local symmetric unitaries, which means they are in the same SPT phase.}.
In what follows, we always implicitly discard qubits that are disentangled when writing down the wave functions.

\subsection{Discrete gauge theories}
\label{sec:discrete}
In this subsection, we remind the reader how to write down the actions for discrete gauge theories. The usual way to implement gauging is to introduce a gauge field $\alpha_1\in H^1(M,\Z_n)$. 
We prefer to use integer cocycles to write an action explicitly.
We assume that our ($1+1$)-manifold is equipped with a triangulation $K$. 
A $\Z_n$ gauge field can be described by an integral cochain $\alpha_1 \in C^1(K, \Z)$ with a constraint $\delta \alpha_1 = n\beta$ , where $\beta$ is an integral 2-cochain.
The constraint can be enforced using a Lagrange multiplier $b \in C^{0}(K, \mathbb{Z})$, and the action including the Lagrange multiplier  is 
\begin{equation}
    S = 2 \pi i \int_M \frac{1}{n} \alpha_1 \cup \delta b.
\end{equation}
This action is the DW theory~\cite{dijkgraaf1990topological}.

The same procedure can be extended to ($2+1$)-manifolds with $k=0 \in H^3(M,\Z_n)$ and we obtain the same action with $b \in C^{1}(K,\mathbb{Z})$:
\begin{equation}
    S = 2 \pi i \int_M \frac{1}{n} \alpha_1 \cup \delta b.
\end{equation}
We can also extend this to any integer $k$, and the action is
\begin{equation}
    S = 2 \pi i \int_M \frac{1}{n} \alpha_1 \cup \delta b + \frac{k}{n^2} \alpha_1 \cup \delta \alpha_1.
\end{equation}
Both actions are invariant under the following shifts in the fields,
\begin{equation}
    \begin{split}
        \alpha_1 \to \alpha_1 + n \gamma,\\
        b \to b + \delta \gamma,
    \end{split}
\end{equation}
where $\gamma \in C^1(K,\Z)$.

\subsection{Generalized cluster SPTs}
In this subsection, we illustrate the outcome of measuring (and post-selecting) a symmetry of a general type of SPT phases and write down the topological actions from the measured states.
The generalized cluster states in $d+1$ dimension is a $\mathbb{Z}_2^{(p)}\times \mathbb{Z}_2^{(d-p-1)}$ SPT state, whose topological action is given by
\beq
   \label{eq:Stop} S^{\text{top}}[M, A_1, A_2]=2\pi i \int_M\frac{1}{2}A_1 \cup A_2,
\eeq
where $A_1$ is $\mathbb{Z}_2$-valued $(p+1)$-cocycle, $A_2$ is $\mathbb{Z}_2$-valued $(d-p)$-cocycle. Upon the replacement $A_1\rightarrow [\delta\phi_1]_2,\ A_2\rightarrow [\delta\phi_2]_2$, the physical action of the SPT order is given by,
\beq
    S=2\pi i \int_M \frac{1}{2}[\delta\phi_1]_2 \cup [\delta\phi_2]_2=2\pi i \int_M \frac{1}{2}\delta\phi_1 \cup \delta\phi_2, 
\label{eq:before_measure}
\eeq
where $\phi_1\in C^p(M,\mathbb{Z}_2)$ and $\phi_1\in C^{d-p-1}(M,\mathbb{Z}_2)$. And the SPT wavefunction on spatial manifold $\partial M$ is given by
\beq                            \ket{SPT}=\sum_{\phi_1,\phi_2}e^{2\pi i \int_{\partial M}\frac{1}{2}\phi_1 \cup \delta\phi_2}\ket{\phi_1,\phi_2}. 
\eeq

To measure the $\phi_2$ field which is charged under the $\mathbb{Z}_2^{(d-p-1)}$ symmetry, the commonly used measurement bases are $Z$ and $X$ (where the latter is associated with the symmetry action). The measurement in $Z$-basis is to measure the local symmetry charge, i.e., the representation of physical field $\phi_2$ on a certain simplex under the $\mathbb{Z}_2^{(d-p-1)}$ symmetry. The measurement in $X$-basis is to measure the local symmetry operator, i.e., the $\mathbb{Z}_2^{(d-p-1)}$ symmetry action on a certain simplex. (We can make a similar statement regarding measuring  the physical $\phi_1$ field,  charged under the $\mathbb{Z}_2^{(p)}$ symmetry.) To make a distinction of two types of measurement, we call the former basis as \emph{``measurement of symmetry charge''}, and the latter basis as \emph{``measurement of symmetry''}~\footnote{In what follows, we will concentrate on the measurement of symmetry, which gives rise to more fruitful results. But we will comment on the measurement of symmetry charge in Sec.~\ref{sec:Z2_cub}, where the measured state possesses computational power.}.

Now suppose we measure (and post-select) the $\mathbb{Z}_2^{(d-p-1)}$ symmetry, i.e., project onto the subspace where $X=1$ on all the $(d-p-1)$-simplices. The measured state excluding disentangled $\phi_2$ degrees of freedom is given by
\beq
    \ket{\psi}=\sum_{\phi_1}\left(\sum_{\phi_2}e^{2\pi i \int_{\partial M}\frac{1}{2}\phi_1 \cup \delta\phi_2}\right)\ket{\phi_1}=\sum_{\phi_1\in Z^p(\partial M,\Z_2)}\ket{\phi_1}.
\eeq

When $p=0$, the cluster state is a $\Z_2\times\Z_2^{(d-1)}$ SPT. The state after measuring the $\Z_2^{(d-1)}$ symmetry is given by
\beq
    \ket{\psi}=\sum_{\phi_1\in Z^0(\partial M,\Z_2)}\ket{\phi_1}=\ket{GHZ}.
\eeq
Therefore, after the measurement, we obtain a  spontaneously $\mathbb{Z}_2$-symmetry breaking phase. When $d=1$ or 2, this result is obtained in the last section using the concrete lattice model.  The ground state subspace of this symmetry-breaking phase in $d+1$ dimension is described by a TFT~\cite{kapustin2014coupling}, 
\beq
S=2\pi i \int_M \frac{1}{2}\delta\phi_1 \cup b,
\eeq
where $b$ is a $\mathbb{Z}$-valued $d$-cochain.

When $p=1$ and $d=2$, from the SPT topological action, the measured state is given by
\beq
    \ket{\psi_{A_2}}=\sum_{\substack{\phi_1\in C^1(M,\mathbb{Z}_2)\\\phi_2\in C^0(M,\mathbb{Z}_2)\\\phi_1|_{\partial M}=\phi}}e^{2\pi i \int_{\partial M} \frac{1}{2}\phi_1\cup \delta\phi_2}\ket{\phi} = \sum_{\phi \in Z^1(\partial M,\Z_2)}\ket{\phi},
\eeq
which is also shown in the last section using the square-lattice example, that, after measuring (and post-selecting) the $\mathbb{Z}_2$ symmetry, we obtain a ground state of the toric code model.

In the general case, the phase after the measurement is described by the topological field with the action given by
\beq
    S=2\pi i \int_M \frac{1}{2}\delta\phi_1\cup b,
    \label{eq:after_measure}
\eeq
where $\phi_1\in C^p(M,\mathbb{Z}_2)$, and $b\in C^{d-p}(M,\mathbb{Z}_2)$. There is an emergent $\mathbb{Z}_2^{(d)}$ symmetry in this action, and the symmetry transformations of $\mathbb{Z}_2^{(p)}\times\mathbb{Z}_2^{(d-p)}$ are given by the following,
\beq
    \phi_1&\rightarrow\phi_1+c^{(p)},\\
    b&\rightarrow b+c^{(d-p)},
\eeq
where $c^{(p)}\in H^p(M,\mathbb{Z}_2)$ and $c^{(d-p)}\in H^{d-p}(M,\mathbb{Z}_2)$.

We can also calculate the anomaly by coupling background gauge fields to the $\mathbb{Z}_2^{(p)}\times\mathbb{Z}_2^{(d-p)}$ symmetry. The action after coupling is given by
\beq
    S=2\pi i \int_M \frac{1}{2}\delta\phi_1\cup b+\frac{1}{2} \phi_1 \cup\Tilde{A}_2+\frac{1}{2} A_1\cup b,
\eeq
where we denote the background $\mathbb{Z}_2^{(d-p)}$ gauge field as $\Tilde{A}_2$, and the background $\mathbb{Z}_2^{(p)}$ gauge field as $A_1$, same as in the topological action in the beginning of this section, i.e., Eq.~(\ref{eq:Stop}), for comparison. This action is not invariant under the gauge transformation $\phi_1\rightarrow \phi_1+\lambda^{(p)}$, $b\rightarrow b+\lambda^{(d-p)}$, $A_1\rightarrow A_1+\delta\lambda^{(p)}$ and $\Tilde{A}_2\rightarrow\Tilde{A}_2+\delta\lambda^{(d-p)}$, and the difference in the actions is
\beq
    \Delta S=2\pi i \int_M \frac{1}{2}\lambda^{(p)}\cup\Tilde{A}_2+\frac{1}{2}A_1\cup\lambda^{(d-p)}+\frac{1}{2}\lambda^{(p)}\cup \delta\lambda^{(d-p)}.
\eeq
 This is the anomaly inflow of an $\mathbb{Z}_2^{(p)}\times \mathbb{Z}_2^{(d-p)}$ SPT in $d+2$ dimensional manifold $X$, such that $\partial X=M$~\cite{tantivasadakarn2021long},
\beq
    S^{\text{top}}=2\pi i \int_X \frac{1}{2}A_1\cup \Tilde{A}_2.
\eeq

We note that the above results can be easily generalized to $\mathbb{Z}_n\times \mathbb{Z}_m$ SPT states if their topological action is of the  same form $S^{\text{top}}\propto A_1\cup A_2$. After measuring the $\mathbb{Z}_n$ (or $\mathbb{Z}_m$) symmetry, the anomalous states are characterized by $\mathcal{L}\propto \Tilde{A}_1\cup A_2$ (or $\mathcal{L}\propto A_1\cup \Tilde{A}_2$).

\subsection{Measurement}
\label{sec:Meas}
In this subsection, we give a procedure for obtaining the actions after measuring a symmetry in topological phases and relating the gauging of a symmetry with measurement.

To conclude and build on the last subsection, starting from an $\mathbb{Z}_2^{(d-p-1)}$ SPT state with physical action (labeled with a subscript `pre')
\beq
    S_{\rm pre}=2\pi i \int_M \frac{1}{2}\delta \phi_1 \cup \delta \phi_2,
\eeq
 measuring the $\mathbb{Z}_2^{(d-p-1)}$ symmetry leads to the measured state is described by the action (labeled with a subscript `m', denoting measurment)
\beq
    S_{\rm m}=2\pi i \int_M \frac{1}{2}\delta \phi_1 \cup b.
    \label{eq:S_m}
\eeq
We can view the measurement (and post-selection) as a map from $S_{\rm pre}$ to $S_{\rm m}$, in which the coboundary $\delta\phi_2$ in $S_{\rm pre}$ is lifted to a cochain $b$ in $S_{\rm m}$. We can pack the analysis into the following diagram:
\begin{equation}
  \begin{tikzpicture}
	\begin{pgfonlayer}{nodelayer}
		\node [style=none] (0) at (-5, -1) {};
		\node [style=none] (1) at (-5, -9) {};
		\node [style=none] (2) at (8.5, -9) {};
		\node [style=none] (3) at (8.5, -1) {};
		\node [style=none, label={above:Partition function}] (4) at (2, 0) {};
		\node [style=none, rotate=270] (5) at (9, -5) {Measuring};
		\node [style=none, rotate=90] (6) at (-5.5, -5) {Lifting};
		\node [style=none] (7) at (-5, -10) {Measured Action};
		\node [style=none] (8) at (-5, 0) {SPT Physical Action };
		\node [style=none] (9) at (-1, 0) {};
		\node [style=none] (10) at (5, 0) {};
		\node [style=none] (11) at (5, -10) {};
		\node [style=none] (12) at (-1, -10) {};
		\node [style=none] (13) at (8.5, -10) {Measured State};
		\node [style=none] (14) at (8.5, 0) {Cluster State};
		\node [style=none, label={above:Partition function}] (15) at (2, -10) {};
	\end{pgfonlayer}
	\begin{pgfonlayer}{edgelayer}
		\draw [style=black] (0.center) to (1.center);
		\draw [style=black] (3.center) to (2.center);
		\draw [style=arrow] (9.center) to (10.center);
		\draw [style=arrow] (12.center) to (11.center);
	\end{pgfonlayer}
\end{tikzpicture}

\end{equation}

From the action after lifting in Eq.~\eqref{eq:S_m}, we can write down the ground states of the topological field theories by calculating partition functions on manifold $M$ with a boundary. We take the toric code action, for example, where $\phi_1$ and $b$ are both $\Z_2$-valued 1-cochains and
the spacetime manifold is a ball $M=B^3$. The ground states $\ket{\psi}$ on $\partial M=S^2$ can be obtained from the path integral formulation of the action, by taking a topological boundary condition $b|_{\partial M}=0$~\cite{chen2023higher},
\beq
    \ket{\psi}=\sum_{\substack{\phi_1,b\in C^1(M,\Z_2),\\\phi_1|_{\partial M}=\phi, \ b|_{\partial M}=0}}e^{S[\phi_1,b]}\ket{\phi}=\sum_{\phi \in Z^1(\partial M,\Z_2)}\ket{\phi},
\eeq
which is exactly the measured state. 

\begin{figure}[h]
\tikzset{every picture/.style={line width=0.75pt}} 
\centering
\resizebox{0.5\linewidth}{!}{\tikzset{every picture/.style={line width=0.75pt}} 

\begin{tikzpicture}[x=0.75pt,y=0.75pt,yscale=-1,xscale=1]

\draw   (70,241.5) .. controls (70,177.43) and (180.59,125.5) .. (317,125.5) .. controls (453.41,125.5) and (564,177.43) .. (564,241.5) .. controls (564,305.57) and (453.41,357.5) .. (317,357.5) .. controls (180.59,357.5) and (70,305.57) .. (70,241.5) -- cycle ;
\draw    (149,220.5) .. controls (175,267.5) and (425,285.5) .. (489,221.5) ;
\draw    (168,236.5) .. controls (195,183.5) and (416,184.5) .. (469,235.5) ;
\draw  [color={rgb, 255:red, 208; green, 2; blue, 27 }  ,draw opacity=0.67 ][line width=1.5]  (110,236) .. controls (110,189.88) and (202.01,152.5) .. (315.5,152.5) .. controls (428.99,152.5) and (521,189.88) .. (521,236) .. controls (521,282.12) and (428.99,319.5) .. (315.5,319.5) .. controls (202.01,319.5) and (110,282.12) .. (110,236) -- cycle ;
\draw [color={rgb, 255:red, 0; green, 0; blue, 0 }  ,draw opacity=0.45 ]   (205,96.5) .. controls (210.46,118.34) and (201.01,151.77) .. (198.52,160.63) ;
\draw [shift={(198,162.5)}, rotate = 284.04] [color={rgb, 255:red, 0; green, 0; blue, 0 }  ,draw opacity=0.45 ][line width=0.75]    (21.86,-6.58) .. controls (13.9,-2.79) and (6.61,-0.6) .. (0,0) .. controls (6.61,0.6) and (13.9,2.79) .. (21.86,6.58)   ;

\draw (157,70.5) node [anchor=north west][inner sep=0.75pt]  [font=\large] [align=left] {defect anyon};

\end{tikzpicture}}
\caption{\label{fig:torus}The ground states on $\partial M$ can be obtained from path integral with insertion of non-contractible defects.}
\end{figure}
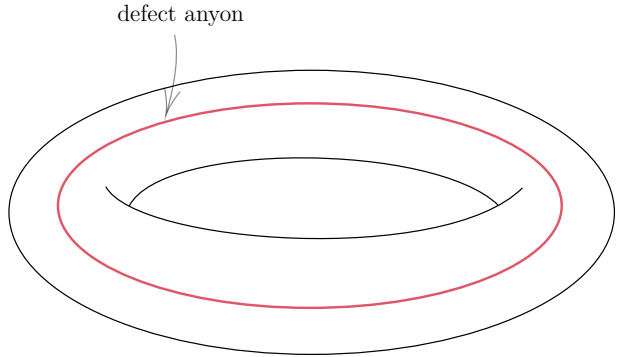

When the spacetime manifold is a solid torus $M=D\times S^1$, the ground state $\ket{\psi_i}$ on $\partial M$ can be obtained from path integral with insertion of non-contractible defects $i$, as in Fig.~\ref{fig:torus}~\cite{bravyi1998quantum,kapustin2011topological}.
In the toric code model, the defect anyons include four different ones: $i=1,e, m,\psi$.
The measured state corresponds to condensing anyon $e$, because the measured state is a superposition of all loops including non-trivial ones. In the defect anyon picture it is equal to gauging the $\Z_2^{(1)}$ 1-form symmetry~\cite{roumpedakis2023higher}. 
The gauging is done by summing over insertions of the $e$-line inside the torus:
\beq
    \ket{\psi}\propto \ket{\psi_1}+ \ket{\psi_e}.
\eeq
This relation between the action and the measured state is denoted in the above diagram by the arrow on the bottom. 

We can generalize this correspondence of lifting in the action to the measurement of symmetry in the topological phase as follows. Suppose in the physical action of a topological phase, we have symmetry $G = \Z_n^{(p)}$ and its corresponding charged field $\phi$, such that the $\Z_n^{(p)}$ symmetry action on the physical fields is 
\beq
\phi\rightarrow \phi +c,
\eeq
where $c$ is a $\Z_n$-valued $p$-cocycle. We claim that by measuring the $\Z_n^{(p)}$ symmetry (and post-selecting), we obtain a phase with a physical action that is given by:
\begin{tcolorbox}
[colback=red!5!white,colframe=red!75!black]
Measuring the $\Z_n^{(p)}$ symmetry of $\phi$ = Lifting $[\delta\phi]_n$ to cochain $b$ in the physical action.
\end{tcolorbox}

We note that the ``measurement/lifting correspondence'' comes with a \emph{caveat}: when the charged field under measured symmetry is not coupled to other fields in the action before measurement, in general, the measured state can not be given by the path integral of the lifted action. For example, for a ($2+1$)d $\Z_2\times G$ SPT, composed from the simple stacking of a $\Z_2$ Levin-Gu SPT and a $G$-SPT states, when measuring the $\Z_2$ symmetry, the corresponding lifted action is the sum of a Chern-Simons action and a physical action for $G$ SPT, as shown below
\beq
    S = 2\pi i \int_M \frac{1}{4}b\cup \delta b +\mathcal{L}[\delta\phi].
\eeq
However, it is known that there is no topological boundary condition for this Chern-Simons action~\cite{kapustin2011topological}. Therefore, the measured state cannot be given from the path integral of the above action. The fact that the lifting does not work in this situation should not bother us too much, because when a measured field $\phi'$ is not coupled to other fields in the action, after measurement, we can simply ignore the disentangled field $\phi'$. In the above example of measuring $\Z_2$ symmetry from the simple stacking of a $\Z_2$ Levin-Gu SPT and a $G$-SPT states, after measurement, we just obtain a $G$-SPT state.

Before ending this section, we have one final remark. From the last subsection, we showed, after measuring a generalized cluster SPT state, we obtain a gauge theory. This relation between the cluster SPT phase and the gauge theory can in fact be generalized as follows. For any theory $\mathcal{T}$ with $\Z_n^{(p)}$ symmetry, where the symmetry action on physical fields is
\beq
\phi\rightarrow \phi + c,
\eeq
where $c$ is a $\Z_n$-valued $p$-cocycle. We have the following~\cite{tantivasadakarn2021long}:
\begin{tcolorbox}[colback=red!5!white,colframe=red!75!black]
Gauging the $\Z_n^{(p)}$ symmetry in $\mathcal{T}$ = Stacking with $\mathcal{T}$ a generalized cluster SPT state and measuring the $\Z_n^{(p)}$ symmetry.
\end{tcolorbox}
The stacking here means adding in the action a cluster SPT physical action $2\pi i\int_M (\frac{1}{n}\delta \phi\cup \delta \phi')$, where $\phi'$ is a $\Z_n$-valued $(d-p-1)$-cocycle. We will discuss the relation between measurement and gauging in more detail in Sec.~\ref{sec:stack cluster} and Sec.~\ref{sec:measure_generalized}.

\section{Measuring 0-form SPTs}
\label{sec:measure_0_form}

\subsection{Stacking of ($1+1$)d cluster states}
\label{sec:stack cluster}
Let us take as example of a $\mathbb{Z}_2^3$ SPT state given by the topological action
\beq
    \label{eq:StopZ23}S^{\text{top}}=2\pi i \int_M \frac{1}{2}A\cup B+\frac{1}{2}B\cup C,
\eeq
which is just a stacking of two cluster states. We can put this SPT state on the $1$d lattice, where qubits of type $a$ and $c$ are on even vertices, and qubits of type $b$ are on odd vertices. The stabilizers of the SPT state are
\beq
    Z^{(b)}_{2k-1}X^{(a)}_{2k}Z^{(b)}_{2k+1},\ Z^{(b)}_{2k-1}X^{(c)}_{2k}Z^{(b)}_{2k+1},\ Z^{(a)}_{2k}Z^{(c)}_{2k}X^{(b)}_{2k+1}Z^{(a)}_{2k+2}Z^{(c)}_{2k+2}.
\eeq
Now we measure the third $\mathbb{Z}_2$ group and post-select the outcome such that $X^{(c)}=1$ uniformly for all type $c$ qubits; then the stabilizers for the measured state are
\beq
    Z^{(b)}_{2k-1}X^{(a)}_{2k}Z^{(b)}_{2k+1},\ Z^{(b)}_{2k-1}Z^{(b)}_{2k+1},\ \prod_{k}X^{(b)}_{2k+1},
\eeq
which gives a GHZ state for type $b$ qubits, and a product state for type $a$ qubits. (Note due to the second stabilizer operator, the first one reduces to $X^{(a)}_{2k}$, showing that type $a$ qubits are not entangled.) 

After type $c$ qubits, we subsequently measure the second $\mathbb{Z}_2$ group (associated with type $b$ qubits) and post-select the outcome such that $X^{(b)}=1$ uniformly for all type $b$ qubits. It is clear that the measured state is of the trivial $\mathbb{Z}_2$ order with stabilizers $X^{(a)}_{2k}=1$.

Equivalently, if we start from the field-theory representation of the SPT state, after measurement of both $\phi_c$ and $\phi_b$ physical fields, the post-measurement state is given by 
\beq
    \ket{\psi}&=\sum_{\phi_a,\phi_b,\phi_c}e^{2\pi i \int_{\partial M}\frac{1}{2}\phi_a\cup \delta\phi_b+\frac{1}{2}\phi_b\cup \delta\phi_c}\ket{\phi_a}\\
    &=\sum_{\phi_a,\phi_b}e^{2\pi i \int_{\partial M}\frac{1}{2}\phi_a\cup \delta\phi_b}\delta_{\delta\phi_b\overset{2}{=}0}\ket{\phi_a}\\
    &=\sum_{\phi_a}\ket{\phi_a}=\ket{+}.
\eeq

We can also follow the procedure described in the last section to directly obtain the TFT description after both type-$c$ and type-$b$ measurements, i.e., 
\beq
    S=2\pi i \int_M \frac{1}{2}\delta\phi_a\cup b +\frac{1}{2}b\cup c.
    \label{eq:stack_cluster}
\eeq
After integrating out fields $b$ and $c$ in the path integral, it is easy to show that the effective action of $\phi_a$ vanishes. We note that there is no emergent symmetry in the above action, and the global symmetry of the measured state is just the remaining $\mathbb{Z}_2$ from measuring $\mathbb{Z}_2^2$ out of $\mathbb{Z}_2^3$. This remaining $\mathbb{Z}_2$ symmetry is anomaly-free. Therefore, we obtain a trivial order after the $\mathbb{Z}_2^2$ measurement. We can also interpret this measurement of both fields as a sequential gauging process. First, by stacking on the $\mathbb{Z}^{(c)}_2$ trivial order a $\mathbb{Z}^{(b)}_2\times\mathbb{Z}^{(c)}_2$ cluster SPT, and measuring the $\mathbb{Z}^{(c)}_2$ symmetry, we gauge the $\mathbb{Z}^{(c)}_2$ trivial order to obtain a $\mathbb{Z}^{(b)}_2$ SSB order, with $\mathbb{Z}^{(b)}_2$ being the emergent ``magnetic'' symmetry after gauging. Next, by stacking on the SSB order a $\mathbb{Z}^{(a)}_2\times\mathbb{Z}^{(b)}_2$ cluster SPT and measuring $\mathbb{Z}^{(b)}_2$ symmetry, we gauge the emergent symmetry in the SSB order to get back to the original $\mathbb{Z}_2$ trivial order. 

\subsection{Another ($1+1$)d $\mathbb{Z}_2^3$ SPT}
Here, we give another example and consider the $\mathbb{Z}_2^3$ SPT order given by the topological action (noting the third term below when comparing to Eq.~(\ref{eq:StopZ23}))
\beq
    S^{\text{top}}=2\pi i \int_M \frac{1}{2}A\cup B+\frac{1}{2}B\cup C+\frac{1}{2}A\cup C.
\eeq
Similar to the previous example, after measuring the second and the third $\mathbb{Z}_2$ groups, the measured state is given by~\footnote{For orientible manifold, the first Stiefel-Whitney class vanishes, then $\phi\cup d\phi=\frac{1}{2}d\left(\phi\cup\phi+\phi\cup_1 d\phi\right)$ is a coboundary, therefore its integral over closed manifold $\partial M$ is zero.}
\beq
    \ket{\psi}&=\sum_{\phi_a,\phi_b,\phi_c}e^{2\pi i \int_{\partial M}\frac{1}{2}\phi_a\cup \delta\phi_b+\frac{1}{2}\phi_b\cup \delta\phi_c+\frac{1}{2}\phi_a\cup \delta\phi_c}\ket{\phi_a}\\
    &=\sum_{\phi_a,\phi_b}e^{2\pi i \int_{\partial M}\frac{1}{2}\phi_a\cup \delta\phi_b}\delta_{\delta\phi_a-\delta\phi_b\overset{2}{=}0}\ket{\phi_a}\\
    &=\sum_{\phi_a}e^{2\pi i \int_{\partial M} \frac{1}{2}\phi_a\cup \delta\phi_a}\ket{\phi_a}\\
    &=\ket{+}.
\eeq

From the lifting procedure, we also know that, after measurement, we obtain a TFT described by
\beq
    S=2\pi i \int_M \frac{1}{2}\delta\phi_a\cup (b+c) +\frac{1}{2}b\cup c\overset{2\pi i }{=}0,
\eeq
which vanishes after integrating out fields $b$ and $c$. We note again the global symmetry of the above action is just the remaining $\mathbb{Z}_2$, with no emergent symmetry.

\subsection{($2+1$)d $\mathbb{Z}_2^3$ SPT (type-3)}
\label{sec:Z2_cub}
\begin{figure}[h]
\includegraphics[width=0.6\linewidth]{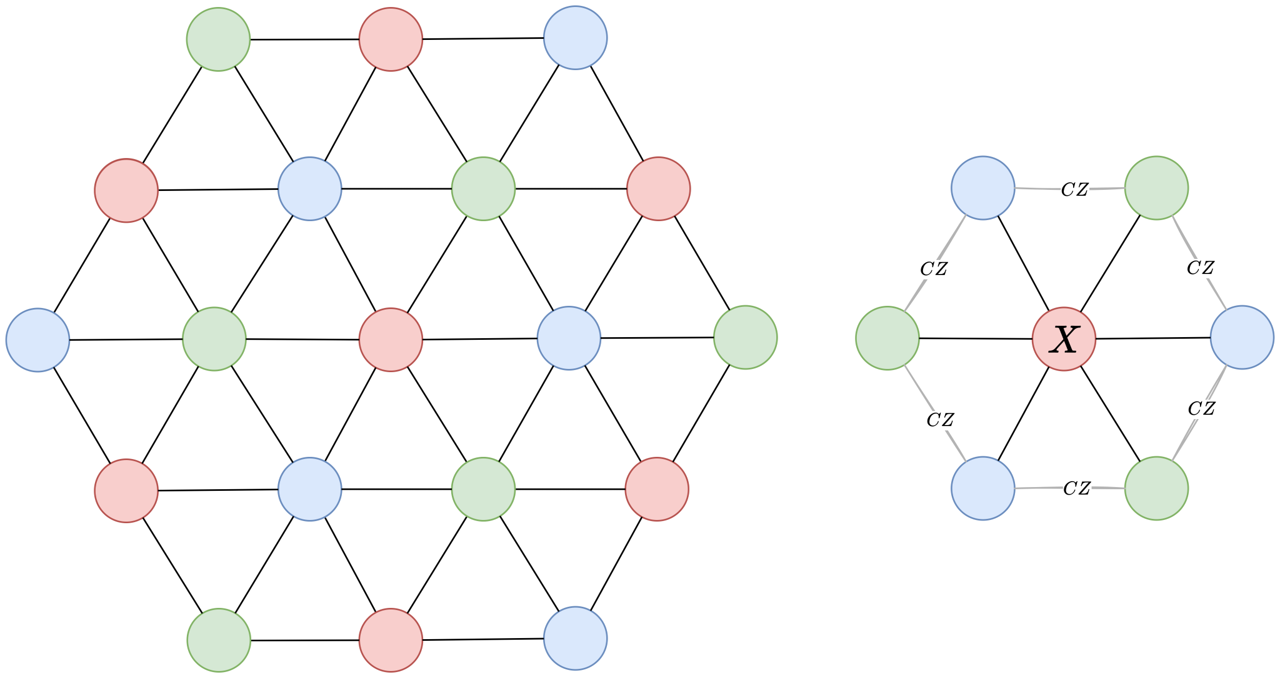}
\caption{\label{fig:2d_Z_2cubic} The three-colorable lattice and the stabilizers for $\mathbb{Z}_2^3$ type-3 SPT states. The $a$, $b$, and $c$ types of sites are colored red, blue, and green, respectively.}
\end{figure}

We now proceed to the case of measuring an SPT state  with only the ordinary symmetry. A nice way to construct the fixed-point wavefunction of $\mathbb{Z}_2^3$ type-3 SPT order is to first place it on a three-colorable lattice and then introduce the stabilizers, composed of $X$ on vertices and $CZ$ on links of the hexagon surrounding it, as illustrated in Fig.~\ref{fig:2d_Z_2cubic}. There are three types of stabilizers depending on the types of the centered sites~\cite{yoshida2016topological}. The Hamiltonian is the sum of all those stabilizers,
\beq
    H=-\sum_{v\in V^{(a)}, V^{(b)},V^{(c)}}S_{v}.
    \label{eq:Z_2_cub_stabilizers}
\eeq

Such a $\Z_2^3$ symmetric SPT state was constructed on a union-jack lattice and shown to be a universal resource for measurement-based quantum computation~\cite{miller2015resource}. One interpretation to understand the quantum computational universality is as follows. By measuring $Z$ operators on a sublattice (i.e., measuring the symmetry charge of one of the $\Z_2$ symmetries), the resultant state without post-selection is a `broken' cluster state with certain edges removed but the connection on average is still above the bond percolation threshold~\cite{wei2018quantum}. Such a broken cluster has sufficient links to support quantum computation by subsequent local measurements. Similarly, a $\Z_n$ generalization on the triangular lattice also gives rise to a resource for quantum computation~\cite{chen2017universal}.

Here, instead of the $\Z_2$ symmetry charge, we measure the third $\mathbb{Z}_2$ symmetry (and post-select $X_{v}=1$ for all the vertices $v\in V^{(c)}$), ignoring the disentangled $c$-qubits, we obtain a state on a hexagonal lattice, respecting $\mathbb{Z}_2^2\times \mathbb{Z}_2^{(1)}$ symmetry. The 0-form symmetry operators are $\prod_{v\in V^{(a)}}X_v$, $\prod_{v\in V^{(b)}}X_v$, and the 1-form symmetry operator is $\prod_{(i,j)\in 1\text{-link}(c)}CZ_{i,j}$, which is shown in Fig.~\ref{fig:2d_hexagon}. In Appendix~\ref{appendix:Z2_sub_measure1}, we show that the measured state is the ground state of a local Hamiltonian respecting the above symmetries,
\beq
H=&-\sum_{v}\left(
\vcenter{\hbox{\includegraphics[scale=.12,trim={0cm 0cm 0cm 0cm},clip]{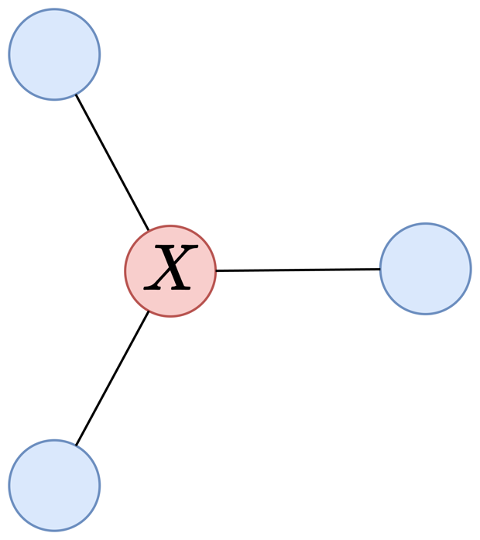}}}
 +\vcenter{\hbox{\includegraphics[scale=.12,trim={0cm 0cm 0cm 0cm},clip]{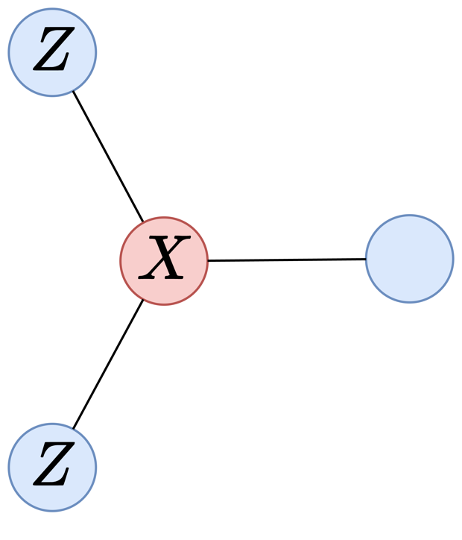}}} + \vcenter{\hbox{\includegraphics[scale=.12,trim={0cm 0cm 0cm 0cm},clip]{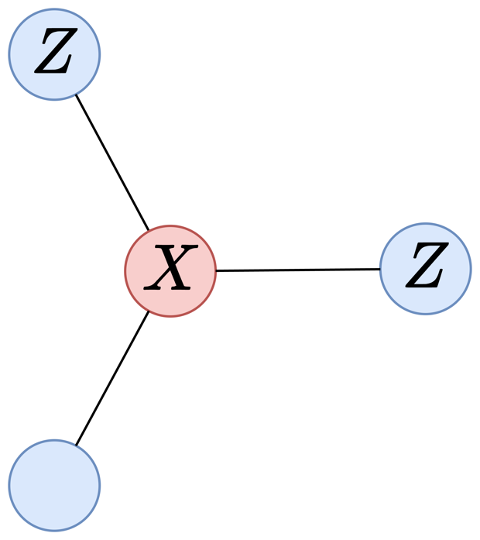}}}\right.  \\
 &\left. + \vcenter{\hbox{\includegraphics[scale=.12,trim={0cm 0cm 0cm 0cm},clip]{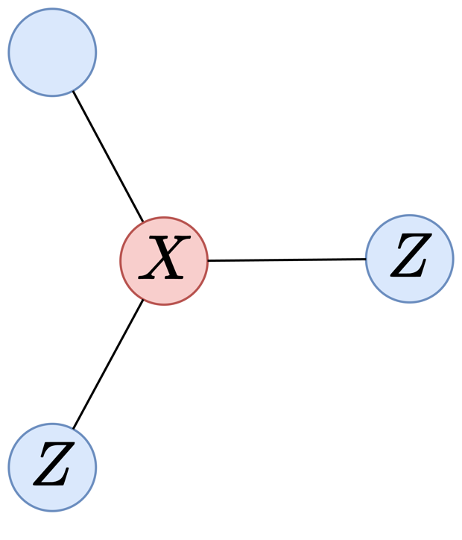}}} -\vcenter{\hbox{\includegraphics[scale=.12,trim={0cm 0cm 0cm 0cm},clip]{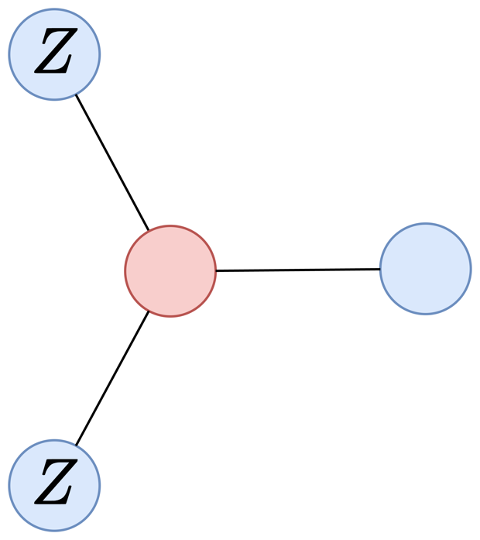}}} - \vcenter{\hbox{\includegraphics[scale=.12,trim={0cm 0cm 0cm 0cm},clip]{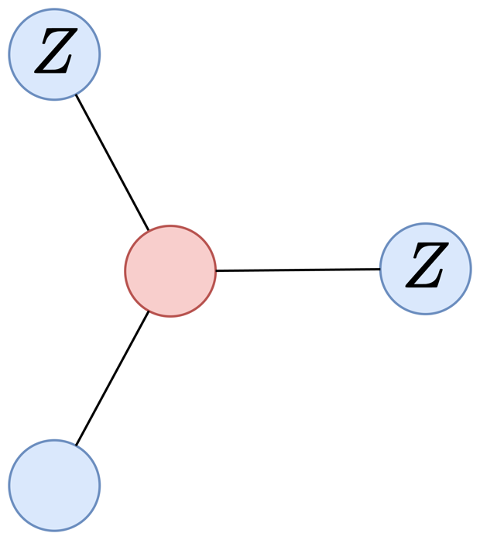}}} - \vcenter{\hbox{\includegraphics[scale=.12,trim={0cm 0cm 0cm 0cm},clip]{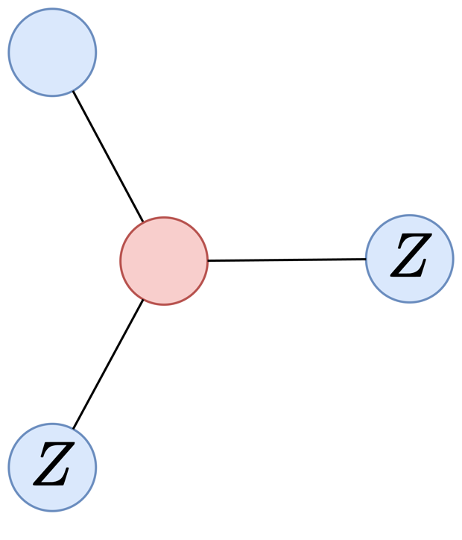}}}\right).
 \label{eq:Hamiltonian1}
\eeq

\begin{figure}[h]
\includegraphics[width=0.6\linewidth]{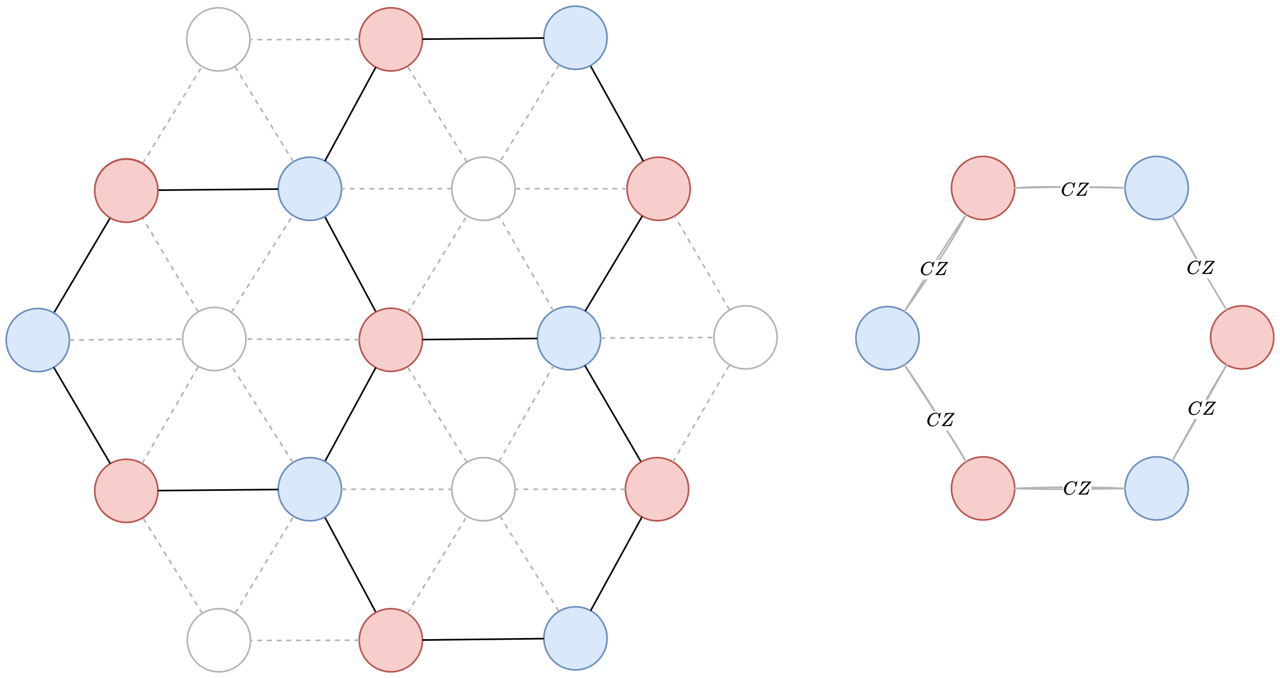}
\caption{\label{fig:2d_hexagon} The hexagonal lattice form by $a$ and $b$ types of sites and the 1-form symmetry in the measured state.}
\end{figure}

To have a field-theory description of the measurement, we start with the topological action of the type-3 SPT phase,
\beq
    S^{\text{top}}=2\pi i \int_M \frac{1}{2}A_1\cup A_2\cup A_3,
\eeq
where $A_1,A_2,A_3$ are $\mathbb{Z}_2$-valued $1$-cocycles. In Appendix~\ref{appendix:spt state}, we show how this topological action gives the lattice model we just introduced. By substituting $A_i\rightarrow [\delta\phi_i]_2$, and lifting $[\delta\phi_3]_2\rightarrow b$ after the measurement, the measured state is described by the TFT action,
\beq
    S=2\pi i \int_M \frac{1}{2}\delta\phi_1\cup \delta\phi_2\cup b,
    \label{eq:phi_phi_b}
\eeq
where $\phi_i$ are $\mathbb{Z}_2$-valued 0-cochains and $b$ is a $\mathbb{Z}_2$-valued 1-cochain. The $\mathbb{Z}_2^2\times\mathbb{Z}_2^{(1)}$ symmetry transformation is given by
\beq
    \phi_1&\rightarrow\phi_1+c_1^{(0)},\\
    \phi_2&\rightarrow\phi_2+c_2^{(0)},\\
    b&\rightarrow b_1+c^{(1)},\\
\eeq
where $c_i^{(0)}$ are $\mathbb{Z}_2$-valued 0-cocycles and $c^{(1)}$ is a $\mathbb{Z}_2$-valued 1-cocycle. The mixed anomaly is characterized by the Lagrangian in one higher dimension: $\mathcal{L}'=\frac{1}{2}A_1\cup A_2\cup \Tilde{A}_3\in H^4(X,\mathbb{Z}_2^2\times\mathbb{Z}_2^{(1)})$. Indeed, the Hamiltonian in Eq.~\eqref{eq:Hamiltonian1} describes a boundary state of a ($3+1$)d SPT state having this topological action, and we can obtain the same measured state from measuring (and post-selecting) the bulk of this SPT state.

\subsection{Further measurements on ($2+1$)d $\mathbb{Z}_2^3$ SPT (type-3)}

\begin{figure}[h]
\includegraphics[width=0.4\linewidth]{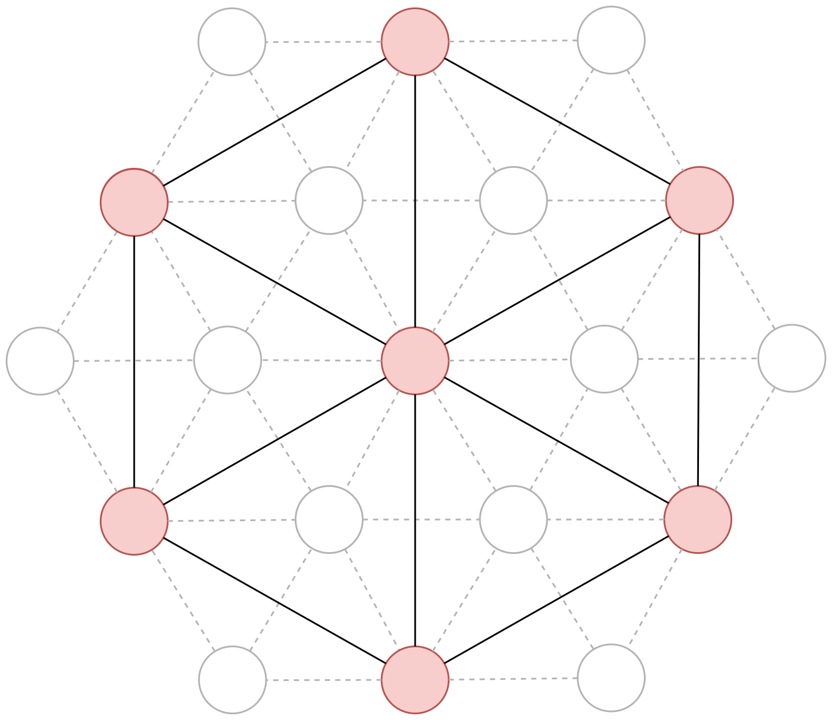}
\caption{\label{fig:2d_tri} The triangular lattice form by $a$ types of sites.}
\end{figure}

One can further perform projective measurements on $X_v=1$ for all $v\in V^{(b)}$, such that the measured state has the remaining degrees of freedom defined on a $V^{(a)}$ triangular lattice as in Fig.~\ref{fig:2d_tri}. 

We prove in Appendix~\ref{appendix:Z2_cub_cont} that the measured state is the ground state of the Hamiltonian,
\beq
    H=\sum_{v\in V^{(a)}}\left(\Phi(\sum_{<v',v>}Z_v Z_{v'})-X_v\right),
    \label{eq:Hamiltonian2}
\eeq
where the function $\Phi(z)=\exp{-\frac{\ln{2}}{2}z+\ln{2}\left(\delta(z-6)-\delta(z+6)\right)}$. The symmetry of this Hamiltonian is generated by the following operators
\beq
    G_{1}=\prod_v X_v,\ G_{2,v}=X_v\Phi^{-1}(\sum_{<v',v>}Z_vZ_{v'}),
\eeq
where $G_1$ generates a global $\mathbb{Z}_2$ symmetry, and $G_{2,v}$ generates a 2-form $\mathbb{Z}_2^{(2)}$ symmetry since $G_{2,v}^2=1$. From the analysis in Ref.~\cite{castelnovo2008quantum} and our Appendix~\ref{appendix:Z2_cub_cont}, this system is in the ``ungauged'' phase of a product state of $\ket{0}_e$ for each edge d.o.f., i.e., this system is in the $\mathbb{Z}_2$ spontaneously breaking phase.

Now we turn to the field-theory description. This state is obtained by measuring one of the $\mathbb{Z}_2$ symmetries in the theory described in Eq.~\eqref{eq:phi_phi_b}. After measurement, $\delta\phi_2$ in the action is lifted to $b'\in C^1(M,\mathbb{Z}_2)$,
\beq
    S'=2\pi i \int_M \frac{1}{2}\delta\phi_1\cup b'\cup b.
\eeq

The global symmetry of this TFT is $\mathbb{Z}_2\times\mathbb{Z}_2^{(2)}$, and the symmetry transformation is given by 
\beq
    \phi_1&\rightarrow\phi_1+c^{(0)},\\
    b'\cup b&\rightarrow b'\cup b+c^{(2)},
\eeq
where $c^{(0)}$ is a $\mathbb{Z}_2$-valued $0$-cocycle and $c^{(2)}$ is a $\mathbb{Z}_2$-valued $2$-cocycle. Coupling the global symmetry to background gauge fields $A_1$ and $\Tilde{B}$, the action is given by
\beq
    S=2\pi i \int_M \frac{1}{2}\delta\phi_1\cup b'\cup b+\frac{1}{2}A_1\cup b'\cup b+\frac{1}{2}\phi_1\cup \Tilde{B}.
\eeq

From the gauge transformation, it can be shown that the $\mathbb{Z}_2\times\mathbb{Z}_2^{(2)}$ anomaly is characterized by $\mathcal{L}'=\frac{1}{2}A_1\cup \Tilde{B}$, which is the same anomaly of a ($2+1$)d GHZ state as we showed previously. This matches the lattice result we obtained above that the measured state is in the $\mathbb{Z}_2$ spontaneously breaking phase. Indeed, the Hamiltonian in Eq.~\eqref{eq:Hamiltonian2} describes a boundary state of a ($3+1$)d SPT state with this topological action, and we can obtain the same measured state from measuring (and post-selecting) the bulk of this SPT state.

\section{Measuring generalized SPTs}
\label{sec:measure_generalized}

\subsection{Another ($2+1$)d $\mathbb{Z}_2\times\mathbb{Z}_2^{(1)}$ SPT}

We have used a $\mathbb{Z}_2\times\mathbb{Z}_2^{(1)}$ SPT (i.e., the cluster state) as a teaser earlier, and now we proceed to consider a more general setting, the result of which naturally extends to $\mathbb{Z}_n\times\mathbb{Z}_n^{(1)}$ SPT orders. Suppose we stack a $\mathbb{Z}_2$ Levin-Gu SPT on the cluster state~\cite{levin2012braiding}, the topological action is given by
\beq
    S^{\text{top}}=2\pi i \int_M \left(\frac{1}{4}A_1\cup \delta A_1+\frac{1}{2}A_1\cup A_2\right).
\eeq
Making replacement $A_i\rightarrow[\delta\phi_i]_2$,
we can obtain the action of this SPT order in terms of physical fields. Then we consider measuring the $\mathbb{Z}_2$ group, which corresponds to lifting $[\delta\phi_1]_2$ to a 1-cochain $b$,
\beq
    S=2\pi i \int_M \left(\frac{1}{4}b\cup \delta b+\frac{1}{2}b\cup \delta\phi_2\right).
    \label{eq:double semion action}
\eeq

From our previous results, we can obtain the anomaly of the emergent $\mathbb{Z}_2^{(1)}\times\mathbb{Z}_2^{(1)}$ symmetry, which is characterized by the topological action in ($3+1$)d
\beq
    S^{\text{top}}=2\pi i \int_X \left(\frac{1}{4}\mathfrak{P}\Tilde{A}_1+\frac{1}{2}\Tilde{A}_1\cup A_2\right),
\eeq
where $\mathfrak{P}\Tilde{A}_1\equiv\Tilde{A}_1 \cup \Tilde{A}_1 - \Tilde{A}_1\cup_1 \delta\Tilde{A}_1$ is the Pontryagin square of $\Tilde{A}_1$~\cite{kapustin2014coupling,tsui2020lattice,chen2023loops}.
This is the stacking of a $\mathbb{Z}_2^{(1)}$ SPT root state and a RBH SPT state~\cite{raussendorf2005long}. This resultant SPT state is known to have the property that the double-semion model can live on its boundary. In fact, as we showed previously, measuring the $\Z_2$ symmetry of Levin-Gu SPT  state stacked with a cluster SPT state is just to gauge the Levin-Gu SPT state, which gives rise to a double-semion model~\cite{tantivasadakarn2021long,bravyi2022adaptive}. We can further gauge the $\Z_2^{(1)}$ symmetry by stacking on the action in Eq.~\eqref{eq:double semion action} another cluster SPT,
\beq
    S=2\pi i \int_M \left(\frac{1}{4}b\cup \delta b + \frac{1}{2}b\cup \delta\phi_2 + \frac{1}{2}\delta \phi \cup \delta\phi_2 \right),
\eeq
where $\phi$ is a $\Z_2$-valued 1-cochain. To gauge the ``magnetic'' $\Z_2^{(1)}$, we measure (and post-select) the symmetry of $\phi_2$, which give rise to action
\beq
    S=2\pi i \int_M \left(\frac{1}{4}b\cup \delta b + \frac{1}{2}b\cup c + \frac{1}{2}\delta \phi \cup c \right),
\eeq
where $c$ is a $\Z_2$-valued 2-cochain. Integrating out field $c$, we get back to the physical action of Levin-Gu SPT state,
\beq
    S=2\pi i \int_M \left(\frac{1}{4}\delta \phi\cup \delta [\delta \phi]_2 \right).
\eeq

\subsection{($2+1$)d $\mathbb{Z}_4\times\mathbb{Z}_4^{(1)}$ SPT}
\label{sec:double-semion}

Now we take a $\mathbb{Z}_4\times\mathbb{Z}_4^{(1)}$ cluster state whose topological action is given below, 
\beq
    S^{\text{top}}=2\pi i \int_M \frac{1}{4}A_1\cup A_2,
\eeq
and consider measuring the $\mathbb{Z}_2$ subgroup of the $\mathbb{Z}_4$ 0-form symmetry. We expect to obtain an SET phase with a $\Z_2$ remaining global symmetry.

We can write down this cluster state on a $2$d square lattice, where qudits (with four levels) live on every vertex and edge. The stabilizers of this SPT state are
\begin{align}
\vcenter{\hbox{\includegraphics[scale=.48,trim={0cm 0cm 0cm 0cm},clip]{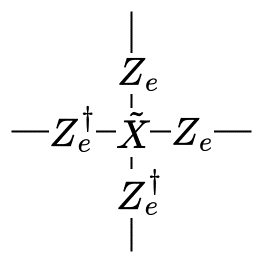}}}, \quad \vcenter{\hbox{\includegraphics[scale=.48,trim={0cm 0cm 0cm 0cm},clip]{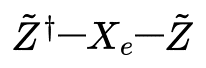}}},
\end{align}
where $\tilde{X}$ and $\tilde{Z}$ are the Pauli operators for qudits on vertices, and $X_e$ and $Z_e$ are Pauli operators for qudits on edges. The Pauli X and Z operators for a four-level qudit are, respectively, defined as 
\beq
    \text{Pauli-}X=\tilde{X}=\sum_{i=0}^3\ket{j+1}\bra{j}, \quad \text{Pauli-}Z=\tilde{Z}=\sum_{i=0}^3\omega^j\ket{j}\bra{j}, 
\eeq
where $\omega=e^{2\pi i/4}$. Measuring the $\mathbb{Z}_2$ subgroup is to project the SPT state onto the subspace where $\tilde{X}^2=1$ for every vertex. After the measurement, we can map each qudit on vertices to a qubit. The Pauli operators of the qubits and the Pauli operators of the original qudits in the subspace have the correspondence: $X \leftrightarrow \tilde{X}$ and $ Z \leftrightarrow \tilde{Z}^2$. After the map, the stabilizers for the measured state are given by
\begin{align}
\vcenter{\hbox{\includegraphics[scale=.48,trim={0cm 0cm 0cm 0cm},clip]{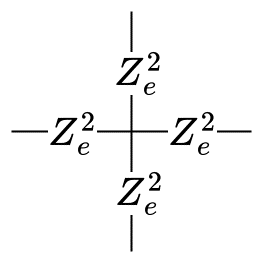}}}, \quad \vcenter{\hbox{\includegraphics[scale=.48,trim={0cm 0cm 0cm 0cm},clip]{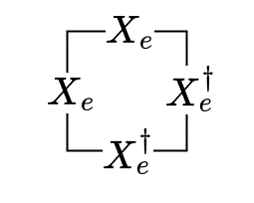}}}, \quad
\vcenter{\hbox{\includegraphics[scale=.48,trim={0cm 0cm 0cm 0cm},clip]{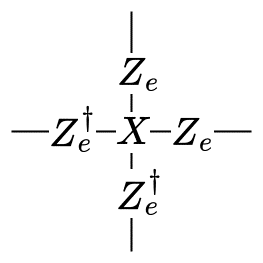}}}, \quad \vcenter{\hbox{\includegraphics[scale=.48,trim={0cm 0cm 0cm 0cm},clip]{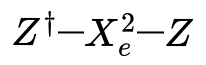}}}.
\end{align}

According to the stabilizers, the measured state has $\mathbb{Z}_2$ 0-form symmetry, $\mathbb{Z}^{(1)}_2$ 1-form symmetry,  and $\mathbb{Z}^{(1)}_4$ 1-form symmetry~\footnote{The corresponding symmetry operators are $\prod_{\text{vertices}}X$, $\prod_{\text{loop on dual lattice}}Z_e^2$ and $\prod_{\text{loop on lattice}}X_e$.}, in which the $\mathbb{Z}^{(1)}_2$ symmetry is spontaneously broken, and the $\mathbb{Z}^{(1)}_4$ symmetry is spontaneously broken to $\mathbb{Z}^{(1)}_2$. We note that the last two terms are stabilizers for the  $\mathbb{Z}_2$ cluster SPT state, despite the degree of freedom on edges being $\mathbb{Z}_4$.\footnote{If we further measure the remaining degrees of freedom on vertices (and post-select  $X=1$), the remaining $\mathbb{Z}^{(1)}_2$ will be broken, and the global symmetry becomes $\mathbb{Z}_4^{(1)}\times\mathbb{Z}_4^{(1)}$. 
The combination of such two-step measurement can be regarded as a one-step measurement of the $\mathbb{Z}_4$ degrees of freedom on the vertices, which leads to a $\mathbb{Z}_4$ toric code, a generalization of the $Z_2$ case. Indeed the measurement can be seen as to gauge the $\mathbb{Z}_4$ trivial order, and the two-step measurement can be seen as to first gauge the $\mathbb{Z}_2$ normal subgroup and then gauge the remaining $\mathbb{Z}_2$ symmetry of the $\mathbb{Z}_4$ trivial order~\cite{li2023symmetry}.

} 
The spontaneous breaking of the 1-form symmetries leads to a topological order. The anyonic excitations of this topological order are created by string operators such as the following,
\begin{align}
\vcenter{\hbox{\includegraphics[scale=.48,trim={0cm 0cm 0cm 0cm},clip]{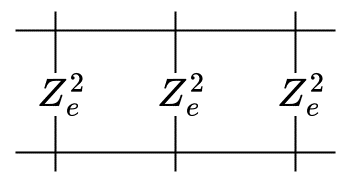}}}, \quad \vcenter{\hbox{\includegraphics[scale=.48,trim={0cm 0cm 0cm 0cm},clip]{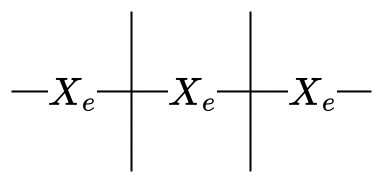}}}.
\end{align}
From the mutual statistics of the anyonic excitations, we find that this topological order is indeed the toric code, while the above two string operators create $m$ and $e$ particles. Further, if we perform the global symmetry within a region $R$ twice, the result is equivalent to the braiding of a $m$ particle along its boundary $\partial R$. This indicates that the measured state is a toric code with a fractionalized $\mathbb{Z}_2$ symmetry. Indeed, this measured state is exactly the state from gauging a $\mathbb{Z}_2$ normal subgroup from a $\mathbb{Z}_4$ trivial order, which is in an SET (symmetry-enriched topological) phase~\cite{barkeshli2019symmetry,li2023symmetry}. 

\begin{figure}[h]
    \centering
    \begin{align}
\left(\vcenter{\hbox{\includegraphics[scale=.36,trim={0cm 0cm 0cm 0cm},clip]{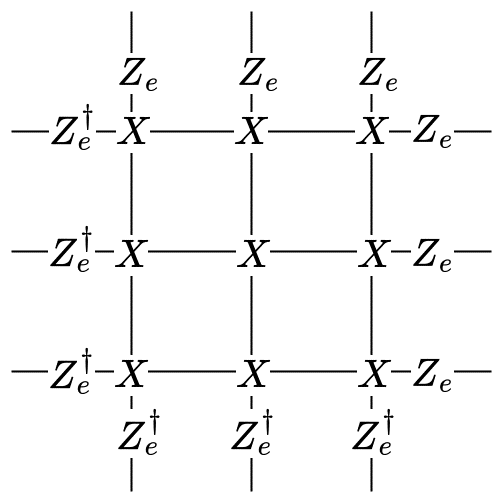}}}\right)^2=\ 
\vcenter{\hbox{\includegraphics[scale=.36,trim={0cm 0cm 0cm 0cm},clip]{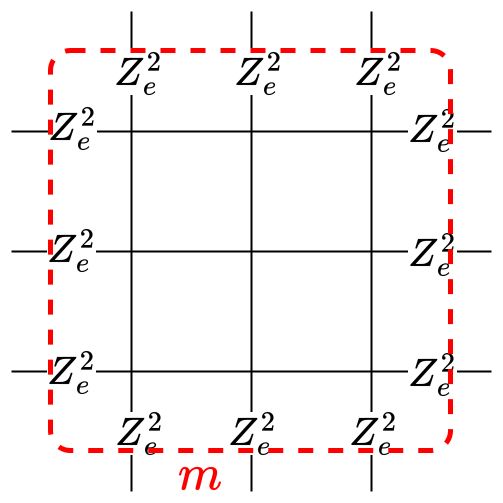}}}.
\end{align}
    \caption{Performing the global symmetry action within a region $R$ twice is equivalent to the braiding of a $m$ particle along its boundary $\partial R$.}
    \label{fig:braiding}
\end{figure}

In the field-theory description, the action in terms of physical fields is $S=2\pi i\int_M \frac{1}{4}\delta\phi_1\cup \delta\phi_2$. To account for the measurement, we make a change of variables,
\beq
    \phi_1=\phi+2\phi',
\eeq
where $\phi:=[\phi_1]_2$ and $\phi':=\frac{1}{2}(\phi_1-[\phi_1]_2)$. Measuring $\tilde{X}^2$ corresponds to measuring the shift operators only on the $\phi'$ field. Therefore, after measurement, $[\delta\phi']_2$ is lifted to $b$. The action becomes,
\beq
    S=2\pi i\int_M \left(\frac{1}{4}\delta\phi\cup \delta\phi_2+\frac{1}{2}b\cup \delta\phi_2\right).
\eeq
We notice that the integral liftings of $\phi$ and $b$ are related. For a different lifting of $\phi$ from $C^0(M,\mathbb{Z}_2)$ to integral-valued, $\tilde{\phi}=\phi+2\lambda$, for a integral-cochain $\lambda$, the cochain $b$ will be shifted by a coboundary, $\tilde{b}=b-\delta\lambda$. Meanwhile, this shift of $b$ is a symmetry of the action. Therefore, the above action is well-defined. The total symmetry transformations of the action are given by,
\beq
    &\mathbb{Z}_2&: \ \phi&\rightarrow \phi+c^{(0)}, &b& \rightarrow  b, &\phi_2&\rightarrow \phi_2,\\
    &\mathbb{Z}^{(1)}_2&: \ \phi&\rightarrow \phi, &b& \rightarrow  b+c^{(1)}_1, &\phi_2&\rightarrow \phi_2,\\
    &\mathbb{Z}^{(1)}_4&: \ \phi&\rightarrow \phi, &b& \rightarrow  b, &\phi_2&\rightarrow \phi_2+c^{(1)}_2.
\label{eq:SET_trans}
\eeq

To observe the symmetry fractionalization, as we showed above, we first ``localize'' the 0-form symmetry action, i.e., to replace the 0-cocycle $c^{(0)}$ by a 0-cochain $\lambda^{(0)}$. The localized symmetry transformation is
\beq
    \phi&\rightarrow \phi+\lambda^{(0)}, &b& \rightarrow  b, &\phi_2&\rightarrow \phi_2.
\eeq 
Applying this localized symmetry transformation twice, we replace $\lambda^{(0)}$ above by $2\lambda^{(0)}$.
However, a different way to achieve the same change in the Lagrangian is by the following transformation,
\beq
    \phi&\rightarrow \phi, &b& \rightarrow  b+\delta\lambda^{(0)}, &\phi_2&\rightarrow \phi_2\\
    \simeq\quad \phi&\rightarrow \phi+2\lambda^{(0)}, &b& \rightarrow  b, &\phi_2&\rightarrow \phi_2.
\eeq 
The first line is exactly a $\mathbb{Z}^{(1)}_2$ 1-form symmetry transformation in the second line of Eq.~\eqref{eq:SET_trans}. This agrees with our lattice analysis above of the symmetry fractionalization in Fig.~\ref{fig:braiding}, where applying the 0-form symmetry action twice in a region is equivalent to inserting an anyon along its boundary.
The anomalous symmetry of this SET phase can be described as a 2-group $\mathbb{G}=(\mathbb{Z}_2,\mathbb{Z}_2\times\mathbb{Z}_4)$, with a nontrivial element from $H^3(\mathbb{Z}_2,\mathbb{Z}_2)$~\cite{kapustin2014anomalies}. The anomaly is characterized by a ($3+1$)d topological action,
\beq
    S^{\text{top}}=2\pi i \int_X \frac{1}{4}\delta A\cup A_2+\frac{1}{2}\Tilde{A}\cup A_2,
\eeq
where $A$ is the $\mathbb{Z}_2$ gauge field, $\Tilde{A}$ is the $\mathbb{Z}_2^{(1)}$ gauge field. The second term in the topological action characterizes the anomaly between the two 1-form symmetries, which corresponds to the toric code. The first term corresponds to a mixed anomaly between the 0-form and 1-form symmetries, which characterizes the symmetry fractionalization class (SFC).

In general, for any theory $\mathcal{T}$ with $\Z_n^{(p)}$ symmetry, where the symmetry action on physical fields is
\beq
\phi\rightarrow \phi +c,
\eeq
where $c$ is a $\Z_n$-valued $p$-cocycle. For a subgroup $\Z_k\subset \Z_n$, we have the following:
\begin{tcolorbox}[colback=red!5!white,colframe=red!75!black]
  Gauging the $\Z_k^{(p)}$-subgroup symmetry in $\mathcal{T}$ = Stacking with $\mathcal{T}$ a $\Z_n^{(p)}$ cluster SPT and measuring the $\Z_k^{(p)}$ symmetry.
\end{tcolorbox}


\subsection{A ($3+1$)d SPT}
In this section, we use a specific model as an example of ($3+1$)d SET orders from measuring SPT orders. We take a $\mathbb{Z}_2\times \mathbb{Z}_2^{(2)}\times\mathbb{Z}_8\times\mathbb{Z}_8^{(2)}\times\mathbb{Z}_2^{(2)}$ SPT state whose topological action is given by
\beq
    S^{\text{top}}=2\pi i \int_M \frac{1}{2}B_1 \cup A_1 +\frac{1}{8}\mathbf{B}_2 \cup \mathbf{A}_2+\frac{1}{2}B_3 \cup A_2+\frac{1}{4}A_1\cup d A_1\cup A_2,
\eeq
where $A_1$ is a $\mathbb{Z}_2$-valued 1-cocycle, $\mathbf{A}_2$ is a $\mathbb{Z}_8$-valued 1-cocycle, $B_1$ is $\mathbb{Z}_2$-valued 3-cocycle, $\mathbf{B}_2$ is $\mathbb{Z}_8$-valued 3-cocycle, and $B_3$ is $\mathbb{Z}_2$-valued 3-cocycle. The action in terms of the physical fields is
\beq
    S=2\pi i \int_M \frac{1}{2}\delta b_1 \cup \delta\phi^a_1 +\frac{1}{8}\delta b_2 \cup \delta{\boldsymbol\phi}^a_2+\frac{1}{2}\delta b \cup \delta{\boldsymbol\phi}^a_2+\frac{1}{4}\delta\phi^a_1\cup \delta[\delta\phi^a_1]_2\cup \delta{\boldsymbol\phi}^a_2,
\eeq
where $b_1$, $\mathbf{b}_2$ and $b$ are the physical fields for $B_1$, $\mathbf{B}_2$ and $B_3$ respectively. Suppose we measure (and post-select) the $\mathbb{Z}_2\times\mathbb{Z}_4$ subgroup of the symmetry. We can make similar changes of variables as the last example 
\beq
    {\boldsymbol\phi}^a_2=\phi+2{\boldsymbol\phi'},\ \text{where}\ \phi=[{\boldsymbol\phi}^a_2]_2, {\boldsymbol\phi'}=\frac{1}{2}({\boldsymbol\phi}^a_2-[{\boldsymbol\phi}^a_2]_2).
\eeq
After measurement, $[\delta\phi^a_1]_2$ is lifted to $a_1$, and $[\delta{\boldsymbol\phi'}]_2$ is lifted to $a_2$. If we further couple the $\mathbb{Z}_2$ 0-form symmetry with background gauge field $A$ and integrate out field $\phi$, the action becomes
\beq
    S=2\pi i \int_M \frac{1}{2}\delta b_1 \cup a_1 +\frac{1}{8}\delta \mathbf{b}_2 \cup A+\frac{1}{4}\delta \mathbf{b}_2 \cup a_2+\frac{1}{2}\delta b\cup A+\frac{1}{4}a_1\cup \delta a_1 \cup A.
\eeq

We note that Ref.~\cite{ye2018three} gives an action for a ($3+1$)d SET order, which is a twisted $\mathbb{Z}_2\times\mathbb{Z}_4$ gauge theory, with a fractionalized 0-form global $\mathbb{Z}_2$ symmetry.  The action we obtain above from measuring SPT is very close to the action in the reference, except for the last term. Ref.~\cite{ye2018three} also proved that their action has an anomaly for the 0-form $\mathbb{Z}_2$ symmetry. Following the same method, we can further prove that the $\mathbb{Z}_2$ symmetry in our action above is anomaly-free.

\section{Measuring topological orders}
\label{sec:measure top order}

\subsection{$\mathbb{Z}_2$ Toric code}
Now, we go beyond SPT phases and turn to the case of measuring ($2+1$)d topological orders. We start with the simplest topological order, i.e., Kitaev's toric code model. The toric code ground state can be obtained from measuring the ($2+1$)d $\mathbb{Z}_2$ cluster SPT state, the action of which is given by
\beq
    S=2\pi i \int_M \frac{1}{2}\delta\phi\cup b,
\eeq
where $\phi$ and $b$ are $\mathbb{Z}_2$-valued 1-cochains. The global symmetry of the toric code is 
\beq
    \phi\rightarrow \phi+c^{(1)}_1,\\
    b\rightarrow b+c^{(1)}_2,
\eeq
where $c^{(1)}_i$ are $\mathbb{Z}_2$-valued 1-cocycles. We showed in previous sections that this global symmetry has an anomaly characterized by $\frac{1}{2}A_1\cup A_2$. If we nevertheless keep on measuring one of the 1-form symmetries (e.g., $\phi$), after measurement, $\delta\phi$ in the action will be lifted to a gauge field $a$, and hence this leads to a TFT action,
\beq
    S=2\pi i \int_M \frac{1}{2}a\cup b,
\eeq
which is trivial because there are no dynamical fields in the action.

On the lattice, the toric code model has star and plaquette stabilizers. After measuring all edges $e$ and post-selecting, $X_e=1$, we thereby obtain a trivial product state. This measurement corresponds to condensing $e$ particles everywhere.

\subsection{$\mathbb{Z}_4$ Toric code}
Now let us consider the  $\mathbb{Z}_4$ toric code, whose  action is given by
\beq
    S=2\pi i \int_M \frac{1}{4}a_1\cup \delta a_2.
\eeq
The global symmetry for this model is $\mathbb{Z}_4^{(1)}\times\mathbb{Z}_4^{(1)}$. A ground state of this order can be given from measuring (and post-selecting) the $\mathbb{Z}_4$ symmetry of a $\mathbb{Z}_4\times\mathbb{Z}_4^{(1)}$ cluster state. The stabilizers of the  $\mathbb{Z}_4$ toric code are 
\begin{align}
\vcenter{\hbox{\includegraphics[scale=.48,trim={0cm 0cm 0cm 0cm},clip]{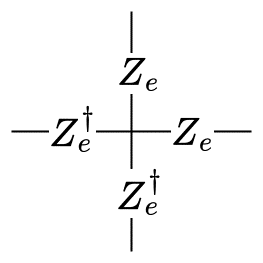}}}, \quad \vcenter{\hbox{\includegraphics[scale=.48,trim={0cm 0cm 0cm 0cm},clip]{figures/Z_4_gauge_face.png}}}.
\end{align}
The global symmetry $\mathbb{Z}_4\times\mathbb{Z}_4^{(1)}$ is spontaneously broken, and the symmetry operators are the closed string operators shown  below, where the first operator there creates $m$ particles and the second operator creates $e$ particles (noting that $m^4=e^4=1$),
\begin{align}
\vcenter{\hbox{\includegraphics[scale=.48,trim={0cm 0cm 0cm 0cm},clip]{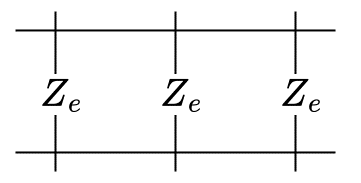}}}, \quad \vcenter{\hbox{\includegraphics[scale=.48,trim={0cm 0cm 0cm 0cm},clip]{figures/Z_4_gauge_e.png}}}.
\end{align}
Measuring the generators of the diagonal $\mathbb{Z}_2^{(1)}$ subgroup will correspond to condensing $e^2m^2$ in the bulk, which gives a double-semion model~\cite{ellison2022pauli}. The stabilizers of the measured state are
\begin{align}
\vcenter{\hbox{\includegraphics[scale=.48,trim={0cm 0cm 0cm 0cm},clip]{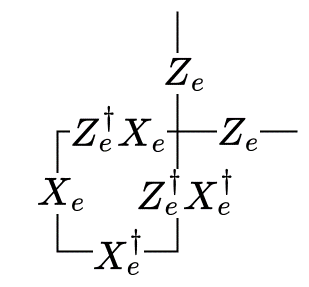}}}, \quad \vcenter{\hbox{\includegraphics[scale=.48,trim={0cm 0cm 0cm 0cm},clip]{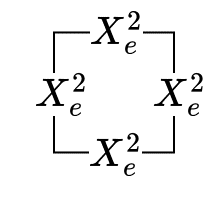}}}, \quad
\vcenter{\hbox{\includegraphics[scale=.48,trim={0cm 0cm 0cm 0cm},clip]{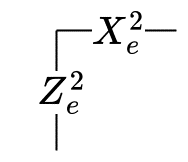}}}, \quad \vcenter{\hbox{\includegraphics[scale=.48,trim={0cm 0cm 0cm 0cm},clip]{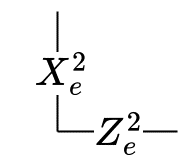}}}.
\end{align}

To obtain the measured phase in the field-theory description, we first make a change of variables,
\beq
    a_3=[a_2]_2,\ a_4=\frac{1}{2}(a_2-[a_2]_2),\ a_5=[a_1+a_2]_2,\ a_6=\frac{1}{2}([a_1+a_2]_4-[a_1+a_2]_2).
\eeq
The action in terms of the new fields is then given as
\beq
    S&=2\pi i \int_M \frac{1}{4}(a_5+2a_6-a_3-2a_4)\cup \delta(a_3+2a_4)\\
    &\overset{1}{=}2\pi i \int_M \frac{1}{4}(a_5-a_3)\cup \delta a_3+\frac{1}{2}a_6\cup \delta a_3-\frac{1}{2}a_4\cup \delta a_3+\frac{1}{2}(a_5-a_3)\cup \delta a_4.
\eeq
After measurement, the field $\delta a_4$ is lifted to $b$, and the action becomes
\beq
    S&=2\pi i \int_M \frac{1}{4}(a_5-a_3)\cup \delta a_3+\frac{1}{2}a_6\cup \delta a_3-\frac{1}{2}b\cup  a_3+\frac{1}{2}(a_5-a_3)\cup b.
\eeq

We now focus on the $\mathbb{Z}_2^{(1)}\times \mathbb{Z}_2^{(1)}$ symmetry and its transformations  are
\beq
    a_3&\rightarrow a_3+c^{(1)}_1, &b& \rightarrow  b-\frac{1}{2}\delta c^{(1)}_1, &a_5&\rightarrow a_5, &a_6&\rightarrow a_6,\\
    a_3&\rightarrow a_3, &b& \rightarrow  b, &a_5&\rightarrow a_5, &a_6&\rightarrow a_6+c^{(1)}_2.
\eeq
Integrating out fields $a_5$ and $b$, we obtain
\beq
    S=2\pi i \int_M \frac{1}{4}a_3\cup \delta a_3 +\frac{1}{2}a_6\cup \delta a_3,
\eeq
which is the action for the double-semion model as shown in Sec.~\ref{sec:double-semion}. Thus, the pictures match.

\section{Conclusion}
\label{sec:conclusion}

In this work, we have presented a field-theoretic framework for describing measurements. We have studied in detail the outcomes of measurements within various topological phases, including the actions and the symmetry anomalies, and demonstrated that these measurements can lead to SPT phases, symmetry spontaneously breaking phases, and topologically ordered phases. This provides a unified framework to predict the post-measurement phases. It is also known that gapped interfaces of topological orders can be created by Kramers-Wannier duality~\cite{yoshida2017gapped} and condensation~\cite{roumpedakis2023higher} implemented by  measurements on the interfaces, and it is possible that our framework can be used to study this. 

We have focused on finite abelian symmetries;
it is, therefore, interesting to extend the current framework to continuous symmetries where the action involves the Wess-Zumino-Witten term~\cite{chen2013symmetry} and nonabelian symmetries for which after measurement noninvertible symmetries are expected to emerge. Notably, the measurements on SPT states with continuous symmetry may also be interpreted as deconfining the Higgs phases~\cite{verresen2022higgs}.

We note that measuring all the symmetry in an SPT phase or measuring the Lagrangian subgroup in a topological order gives rise to the strange correlator~\cite{you2014wave,lepori2023strange}, which often shows some long-ranged behavior.
It would also be intriguing to probe the behavior of the strange correlators and their possible relations to the pseudo entropy using the field-theoretic framework proposed here~\cite{doi2023pseudoentropy}. This will be an interesting direction to  explore.

Measurement-based quantum computation employs resource states~\cite{briegel2009measurement,wei2018quantum}, such as those in the SPT phases, and performs local measurements to induce computation. There, however, the measurement basis may need to be changed. In our case, the basis for measurement is fixed. It would be interesting to extend our formalism to deal with measurements not just the local operators of the  symmetry, nor the local symmetry charge, but to a more general basis. This may potentially allow us to use a field-theoretical language to discuss measurement-induced phase transitions~\cite{li2018quantum,skinner2019measurement,li2019measurement}.  We hope this work offers some insight into understanding measurements within topological phases and their applications in quantum information processing.  
\begin{acknowledgments}
Y. L. would like to thank Jiahao Hu, Ruochen Ma, and Hiroki Sukeno for useful discussions. M. L. appreciates remarks from Yichil Choi.
 This work was partly supported by the National Science Foundation under Grant No. PHY 2310614. T.-C.W. also acknowledges the support by Stony Brook University's Center for Distributed Quantum Processing. 
\end{acknowledgments}

\bibliography{ref}

\begin{widetext}

\appendix

\section{Type-3 SPT state from topological action}
\label{appendix:spt state}
The topological action of a ($2+1$)d $\mathbb{Z}_2^3$ type-3 SPT is given by,
\beq
    S^{\text{top}}=2\pi i \int_M \frac{1}{2}A_1\cup A_2\cup A_3,
\eeq
where $A_1,A_2,A_3$ are $\mathbb{Z}_2$-valued $1$-cocycles. To write down an SPT state from this topological action, we replace the background gauge fields by physical fields $A_i\rightarrow [\delta\phi_i]_2$, and put the SPT order on a triangulated 3-colorable lattice. The SPT state is given by
\beq
    \ket{SPT}=\sum_{\phi_i}e^{2\pi i \int_M \frac{1}{2}\phi_1\cup \delta\phi_2 \cup \delta\phi_3}\ket{\phi_1,\phi_2,\phi_3}.
\eeq
The branching structure of a lattice is to assign the ordering of vertices; on this colorable lattice, we choose the branching structure to be $a\leftarrow b\leftarrow c$. We can thus write the wavefunction amplitude as $(-1)^{\chi(\phi_1, \phi_2, \phi_3)}$, where
\beq
\chi(\phi_1, \phi_2, \phi_3)&=\sum_{\Delta{abc}}\phi_1(a)(\phi_2(a)+\phi_2(b))(\phi_3(b)+\phi_3(c))\\
&=\sum_{\Delta{abc}}\phi_1(a)\phi_2(a)\phi_3(b)+\phi_1(a)\phi_2(a)\phi_3(c)+\phi_1(a)\phi_2(b)\phi_3(b)+\phi_1(a)\phi_2(b)\phi_3(c)\\
&=\sum_{\Delta{abc}}\phi_1(a)\phi_2(b)\phi_3(c),
\eeq
where the first three terms in the summation vanish because they are supported only on edges, and contributions from adjacent triangles cancel. Therefore, the wavefunction amplitude depends only on the $\phi_1$ fields on $a$-type sites, $\phi_2$ fields on $b$-type sites, and $\phi_3$ field on $c$-type sites. To study the SPT state, we can ignore all the other disentangled qubits. Therefore, we have just one qubit per site, and the wavefunction amplitude $(-1)^{\chi(\phi_1, \phi_2, \phi_3)}=\prod_{\Delta{abc}}CCZ_{a,b,c}$. The stabilizers of this SPT state are exactly given by Eq.~\eqref{eq:Z_2_cub_stabilizers}.

\section{The measured states of type-3 SPT}
\label{appendix:Z2_sub_measure1}
After we measure the $\mathbb{Z}_2^{(c)}$ symmetry (and post-select $X_v=1$ for all $v\in V^{(c)}$), the measured state is given by
\beq
    \ket{\psi}=\left(\prod_{v\in V^{(c)}} \bra{+}\right)\left(\prod_{\Delta_{abc}} CCZ_{a,b,c}\right)\left(\prod_{v\in V^{(a)},V^{(b)},V^{(c)}}\ket{+}\right).
\eeq
We for now focus on one site $v$ of type-$a$ as below, where we denote $a$-type sites as red, and $b$-type sites as blue.
\begin{align}
\vcenter{\hbox{\includegraphics[scale=.24,trim={0cm 0cm 0cm 0cm},clip]{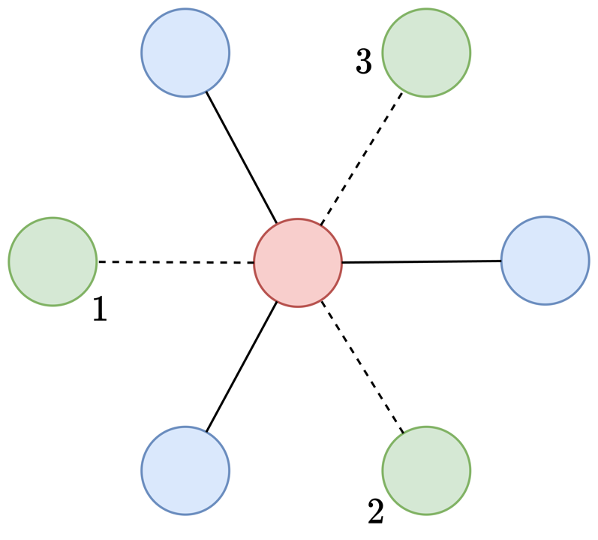}}}
\end{align}
We denote the three $c$-sites adjacent to $v$ as $c_1, c_2$ and $c_3$, and define 
\beq
    \ket{\psi_{c1,c2,c3}}:=\bra{c_1,c_2,c_3}\prod_{v'\in V^{(c)}\backslash \{c_1,c_2,c_3\}}\bra{+}_{v'}\ket{SPT}.
\eeq
Then the measured state $\ket{\psi}$ can be written as 
\beq
    \ket{\psi}=\frac{1}{2\sqrt{2}}\sum_{c_1,c_2,c_3=0,1}\ket{\psi_{c1,c2,c3}}.
\eeq

Recalling that the stabilizers of the SPT state are given by
\begin{align}
\vcenter{\hbox{\includegraphics[scale=.24,trim={0cm 0cm 0cm 0cm},clip]{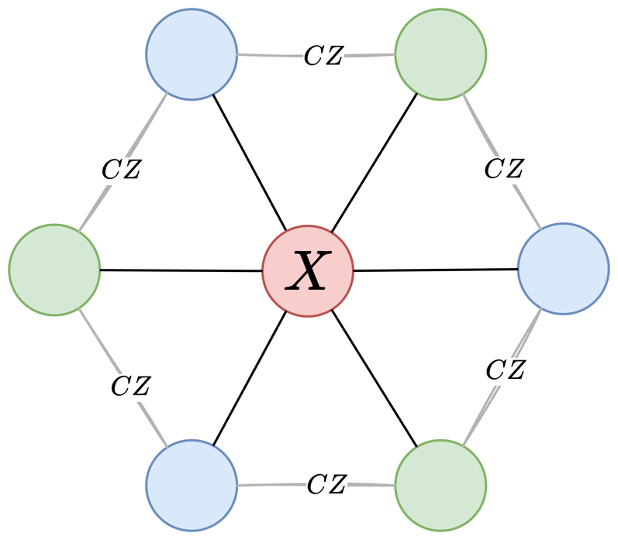}}},
\end{align}
we can calculate 
\beq
    &\left(
\vcenter{\hbox{\includegraphics[scale=.12,trim={0cm 0cm 0cm 0cm},clip]{figures/Z2_cub_x.png}}}
 + \vcenter{\hbox{\includegraphics[scale=.12,trim={0cm 0cm 0cm 0cm},clip]{figures/Z2_cub_zxz1.png}}} + \vcenter{\hbox{\includegraphics[scale=.12,trim={0cm 0cm 0cm 0cm},clip]{figures/Z2_cub_zxz2.png}}} + \vcenter{\hbox{\includegraphics[scale=.12,trim={0cm 0cm 0cm 0cm},clip]{figures/Z2_cub_zxz3.png}}}\right)\ket{\psi_{1,0,0}}\\
 &=\left(
...\right)\bra{1,0,0}\prod_{v'}\bra{+}_{v'}\ket{SPT}\\
 &=\left(
...\right)\bra{1,0,0}\prod_{v'}\bra{+}\vcenter{\hbox{\includegraphics[scale=.12,trim={0cm 0cm 0cm 0cm},clip]{figures/Z2_cub_stab.png}}}\ket{SPT}\\
&=\left(
...\right)\vcenter{\hbox{\includegraphics[scale=.12,trim={0cm 0cm 0cm 0cm},clip]{figures/Z2_cub_zxz1.png}}}\bra{1,0,0}\prod_{v'}\bra{+}\ket{SPT}\\
&=\left(
\vcenter{\hbox{\includegraphics[scale=.12,trim={0cm 0cm 0cm 0cm},clip]{figures/Z2_cub_x.png}}}
 + \vcenter{\hbox{\includegraphics[scale=.12,trim={0cm 0cm 0cm 0cm},clip]{figures/Z2_cub_zxz1.png}}} + \vcenter{\hbox{\includegraphics[scale=.12,trim={0cm 0cm 0cm 0cm},clip]{figures/Z2_cub_zxz2.png}}} + \vcenter{\hbox{\includegraphics[scale=.12,trim={0cm 0cm 0cm 0cm},clip]{figures/Z2_cub_zxz3.png}}}\right)\vcenter{\hbox{\includegraphics[scale=.12,trim={0cm 0cm 0cm 0cm},clip]{figures/Z2_cub_zxz1.png}}}\ket{\psi_{1,0,0}}\\
 &=\left(1 + \vcenter{\hbox{\includegraphics[scale=.12,trim={0cm 0cm 0cm 0cm},clip]{figures/Z2_cub_zz1.png}}} + \vcenter{\hbox{\includegraphics[scale=.12,trim={0cm 0cm 0cm 0cm},clip]{figures/Z2_cub_zz2.png}}} + \vcenter{\hbox{\includegraphics[scale=.12,trim={0cm 0cm 0cm 0cm},clip]{figures/Z2_cub_zz3.png}}}\right)\ket{\psi_{1,0,0}}.
\eeq

Repeat the calculation for each $\ket{\psi_{c_1,c_2,c_3}}$, it is then straightforward to verify that 
\beq
    &\left(
\vcenter{\hbox{\includegraphics[scale=.12,trim={0cm 0cm 0cm 0cm},clip]{figures/Z2_cub_x.png}}}
 + \vcenter{\hbox{\includegraphics[scale=.12,trim={0cm 0cm 0cm 0cm},clip]{figures/Z2_cub_zxz1.png}}} + \vcenter{\hbox{\includegraphics[scale=.12,trim={0cm 0cm 0cm 0cm},clip]{figures/Z2_cub_zxz2.png}}} + \vcenter{\hbox{\includegraphics[scale=.12,trim={0cm 0cm 0cm 0cm},clip]{figures/Z2_cub_zxz3.png}}}\right)\ket{\psi}\\
 &=\left(1 + \vcenter{\hbox{\includegraphics[scale=.12,trim={0cm 0cm 0cm 0cm},clip]{figures/Z2_cub_zz1.png}}} + \vcenter{\hbox{\includegraphics[scale=.12,trim={0cm 0cm 0cm 0cm},clip]{figures/Z2_cub_zz2.png}}} + \vcenter{\hbox{\includegraphics[scale=.12,trim={0cm 0cm 0cm 0cm},clip]{figures/Z2_cub_zz3.png}}}\right)\ket{\psi}.
\eeq

We define an operator $\mathcal{O}_v$ supported locally around site $v$ as,
\beq
    \mathcal{O}_v:=&-
\vcenter{\hbox{\includegraphics[scale=.12,trim={0cm 0cm 0cm 0cm},clip]{figures/Z2_cub_x.png}}}
 - \vcenter{\hbox{\includegraphics[scale=.12,trim={0cm 0cm 0cm 0cm},clip]{figures/Z2_cub_zxz1.png}}} - \vcenter{\hbox{\includegraphics[scale=.12,trim={0cm 0cm 0cm 0cm},clip]{figures/Z2_cub_zxz2.png}}} - \vcenter{\hbox{\includegraphics[scale=.12,trim={0cm 0cm 0cm 0cm},clip]{figures/Z2_cub_zxz3.png}}} \\
 & +\vcenter{\hbox{\includegraphics[scale=.12,trim={0cm 0cm 0cm 0cm},clip]{figures/Z2_cub_zz1.png}}} + \vcenter{\hbox{\includegraphics[scale=.12,trim={0cm 0cm 0cm 0cm},clip]{figures/Z2_cub_zz2.png}}} + \vcenter{\hbox{\includegraphics[scale=.12,trim={0cm 0cm 0cm 0cm},clip]{figures/Z2_cub_zz3.png}}}.
\eeq
From the above analysis, the measured state $\ket{\psi}$ is an eigenstate of $\mathcal{O}_v$ with eigenvalue $-1$, for every vertex $v$. It is easy to verify that this operator satisfies the relation, $\mathcal{O}_v^2-6\mathcal{O}_v-7=0$; thus the lowest eigenvalue of $\mathcal{O}_v$ is $-1$. Therefore, we conclude that the measured state is the ground state of Hamiltonian,
\beq
    H=\sum_{v\in V^{(a)},V^{(b)}}\mathcal{O}_v.
\eeq

\section{The measured states of the type-3 SPT order (cont.)}
\label{appendix:Z2_cub_cont}
After we measure the $\mathbb{Z}^{(b)}_2\times\mathbb{Z}_2^{(c)}$ symmetry (and post-select $X_v=1$ for all $v\in V^{(b)}, V^{(c)}$), the measured state is given by
\beq
    \ket{\psi'}=\left(\prod_{v\in V^{(b)}, V^{(c)}} \bra{+}\right)\left(\prod_{\Delta_{abc}} CCZ_{a,b,c}\right)\left(\prod_{v\in V^{(a)},V^{(b)},V^{(c)}}\ket{+}\right).
\eeq
The $a$-type qubits form a triangular lattice, 
\begin{align}
\vcenter{\hbox{\includegraphics[scale=.24,trim={0cm 0cm 0cm 0cm},clip]{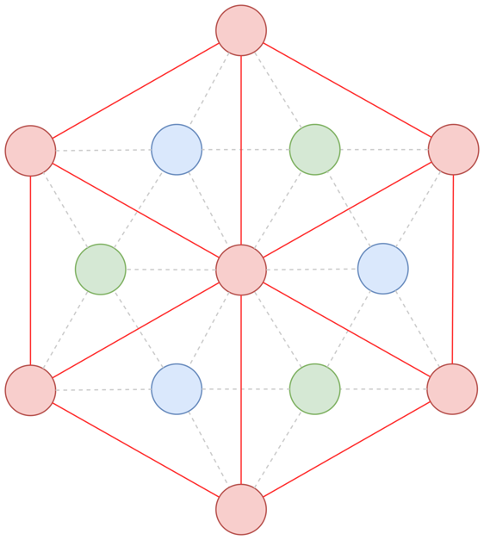}}}.
\end{align}
Again, we first focus on one site $v$ of type-$a$. Using a similar method as in our last calculation, we obtain that
\beq
    X_v\ket{\psi'}=\Phi(\sum_{<v',v>}Z_v Z_{v'})\ket{\psi'},
\eeq
where the function $\Phi(z)=\exp{-\frac{\ln{2}}{2}z+\ln{2}\left(\delta(z-6)-\delta(z+6)\right)}$. The physical meaning of this function is to count the number of domain wall between site $v$ and 6 adjacent $a$-type sites,
\beq
    \Phi(\sum_{<v',v>}Z_v Z_{v'})=\begin{cases}
4,\quad \text{number of d.w.}=0,1, \\
2,\quad \text{number of d.w.}=2,\\
1,\quad \text{number of d.w.}=3, \\
\frac{1}{2},\quad \text{number of d.w.}=4, \\
\frac{1}{4},\quad \text{number of d.w.}=5,6. \\
\end{cases} 
\eeq
We denote $\Phi(\sum_{<v',v>}Z_v Z_{v'})$ as $\Phi_v$, the operator $(\Phi_v-X_v)$ has algebra,
\beq
    (\Phi_v-X_v)(\Phi_v-X_v)&=\Phi_v^2-\Phi_v X_v - X_v \Phi_v + 1 \\
    &=\Phi_v^2- X_v \Phi_v^{-1} - X_v \Phi_v + 1\\
    &=(\Phi_v-X_v)(\Phi_v +\Phi_v^{-1})\\
    &=2 (\Phi_v-X_v)\cosh{\Phi_v}.
\eeq
Since operator $\cosh{\Phi_v}>0$, and $[\cosh{\Phi_v},(\Phi_v-X_v)]=0$, we have $(\Phi_v-X_v)\geq0$. Therefore, we conclude that the measured state is the ground state of Hamiltonian,
\beq
    H=\sum_v \Phi_v-X_v.
\eeq    

This Hamiltonian has a global $\mathbb{Z}_2$ symmetry, therefore we can gauge the symmetry (also known as the Kamers-Wannier duality) to obtain a lattice model where qubits live on the edges of the triangular lattice. The Hamiltonian of the gauged model is given by
\beq
    H=-\sum_{v} \left(\prod_{e\supset v}X_e\right) +\sum_v \Phi(\sum_{e\supset v}Z_e) -\sum_{f}\left(\prod_{e\subset f}Z_e\right),
\eeq
where $v$ is a vertex on the lattice, and $f$ is a face on the lattice. The first term and the third term are exactly the star term and the plaquette term in the toric code model. If the second term vanishes, a ground state of this Hamiltonian is a toric code ground state, which is a equal-weighted sum over all closed-loop configurations. If the second term is present, the ground state becomes a weighted sum over all closed-loop configurations. When the weight is close to $1$, the model is still in the topological order. When the weight is far from $1$, the model eventually has a ground state where $Z_e=1$ for all the edges. This phase can be understood as condensing all the $m$ particles in the bulk. From Ref.~\cite{castelnovo2008quantum}, when the weight is of the form $\Phi(\sum_{e\supset v}Z_e)=e^{-\beta \sum_{e\supset v}Z_e}$, the critical value of the summing weight to experience a phase transition out of the topological order is the same as the critical value of reduced nearest-neighbor coupling $J/T=\beta$ in a $2$d classical Ising model. On a triangular lattice, the critical coupling constant of Ising model is $\beta_c\approx 0.22$, which is smaller than the conservative estimate of our coupling constant $\ln{2}/3$. Therefore, we conclude that the gauged model is in the phase where $m$ particle is condensed in the entire bulk, which means that the original lattice model we write down is in a $\mathbb{Z}_2$ SSB phase.

\end{widetext}
\end{document}